\documentclass[english,american]{article}
\usepackage[latin9]{inputenc}
\usepackage[a4paper]{geometry}
\usepackage{color}
\usepackage{babel}
\usepackage{prettyref}
\usepackage{amsmath}
\usepackage{amssymb}
\usepackage{graphicx}
\usepackage{setspace}
\usepackage[unicode=true,
 bookmarks=true,bookmarksnumbered=false,bookmarksopen=false,
 breaklinks=false,pdfborder={0 0 1},backref=false,colorlinks=false]
 {hyperref}
\hypersetup{pdftitle={Correlation Structure is intrinsic},
 pdfauthor={Helias, Tetzlaff, Diesmann},
 pdfsubject={correlated neural networks},
 colorlinks=true,linkcolor=blue,citecolor=blue,urlcolor=blue,filecolor=blue}
\usepackage{breakurl}

\makeatletter
\newrefformat{cap}{\hyperref[#1]{Figure~\ref{#1}}}
\newrefformat{fig}{\hyperref[#1]{Figure~\ref{#1}}}
\newrefformat{tab}{\hyperref[#1]{Table~\ref{#1}}}
\newrefformat{sec}{\hyperref[#1]{Sec.~\ref{#1}}}
\newrefformat{sub}{\hyperref[#1]{Sec.~\ref{#1}}}
\newrefformat{alg}{\hyperref[#1]{Alg.~\ref{#1}}}
\newrefformat{app}{\hyperref[#1]{App.~\ref{#1}}}

\usepackage[OT1]{fontenc}

\renewenvironment{abstract}
{\noindent{\normalfont\large\textbf{Abstract}}\par\vspace{0.5\baselineskip}}
{\par}

\renewcommand{\@seccntformat}[1]{%
\csname the#1\endcsname\hspace{0.5em}}

\renewcommand{\section}{\@startsection
{section}%
{1}%
{0mm}%
{-\baselineskip}%
{0.5\baselineskip}%
{\normalfont\large\bfseries}}

\renewcommand{\subsection}[1]{\ssubsection{#1.}}
\newcommand{\ssubsection}{\@startsection
{subsection}%
{2}%
{1em}%
{-\baselineskip}%
{-\fontdimen2\font plus -\fontdimen3\font minus -\fontdimen4\font}%
{\normalfont\bfseries}}

\renewcommand{\subsubsection}[1]{\sssubsection{#1.}}
\newcommand{\sssubsection}{\@startsection
{subsubsection}%
{3}%
{1em}%
{-\baselineskip}%
{-\fontdimen2\font plus -\fontdimen3\font minus -\fontdimen4\font}%
{\normalfont\itshape}}

\usepackage{footmisc}

\renewcommand{\@makecaption}[2]{%
{\parbox[t]{\linewidth}{%
\normalsize\renewcommand{\baselinestretch}{1.0}\normalsize
\vspace{2mm}
\textbf{#1:} #2
}}}

\usepackage{xspace}

\usepackage{xcolor}
\definecolor{lightgray}{gray}{0.9}

\usepackage{listings}
\makeatletter
\renewcommand\@biblabel[1]{#1.}
\makeatother

\makeatother

\begin{document}
\global\long\def\n{\mathbf{n}}
\global\long\def\u{\mathbf{u}}
\global\long\def\W{\mathbf{W}}
\global\long\def\erf{\mathrm{erf}}
\global\long\def\erfc{\mathrm{erfc}}
\global\long\def\eig{\mathrm{eig}}
\global\long\def\second{\,\mathrm{s}}
\global\long\def\ms{\,\mathrm{ms}}
\let\oldnameref\nameref \renewcommand{\nameref}[1]{\textbf{``\oldnameref{#1}''}}

\begin{titlepage}\thispagestyle{empty}\setcounter{page}{0}\pdfbookmark[1]{Title}{TitlePage}

\begin{center}
\textbf{\textcolor{black}{\large The correlation structure of local
neuronal networks intrinsically results from recurrent dynamics}}
\par\end{center}{\large \par}

\begin{center}
\textbf{Moritz Helias$^{1}$, Tom Tetzlaff$^{1}$, Markus Diesmann$^{1,2}$}
\par\end{center}

\vspace{3cm}
$^{1}$\parbox[t]{12cm}{Institute of Neuroscience and Medicine (INM-6)
and Institute for Advanced Simulation (IAS-6), Jülich Research Centre
and JARA, Jülich, Germany}\\[3mm]

$^{2}$\parbox[t]{12cm}{Medical Faculty, RWTH Aachen University,
Aachen, Germany}\\[3mm]

\noindent\raisebox{11.8cm}[0cm][0cm]{\hspace*{12.5cm}\textbf{}}\vfill{}
\noindent\\
Correspondence to:\hspace{1em}\parbox[t]{11cm}{Dr.\ Moritz Helias\\

\newlength{\myw}\settowidth{\myw}{fax:\ }\makebox[\myw][l]{tel: +49-2461-61-9467}

\href{mailto:m.helias@fz-juelich.de}{m.helias@fz-juelich.de}

}\end{titlepage}
\begin{abstract}
\noindent Correlated neuronal activity is a natural consequence of
network connectivity and shared inputs to pairs of neurons, but the
task-dependent modulation of correlations in relation to behavior
also hints at a functional role. Correlations influence the gain of
postsynaptic neurons, the amount of information encoded in the population
activity and decoded by readout neurons, and synaptic plasticity.
Further, it affects the power and spatial reach of extracellular signals
like the local-field potential. A theory of correlated neuronal activity
accounting for recurrent connectivity as well as fluctuating external
sources is currently lacking. In particular, it is unclear how the
recently found mechanism of active decorrelation by negative feedback
on the population level affects the network response to externally
applied correlated stimuli. Here, we present such an extension of
the theory of correlations in stochastic binary networks. We show
that (1) for homogeneous external input, the structure of correlations
is mainly determined by the local recurrent connectivity, (2) homogeneous
external inputs provide an additive, unspecific contribution to the
correlations, (3) inhibitory feedback effectively decorrelates neuronal
activity, even if neurons receive identical external inputs, and (4)
identical synaptic input statistics to excitatory and to inhibitory
cells increases intrinsically generated fluctuations and pairwise
correlations. We further demonstrate how the accuracy of mean-field
predictions can be improved by self-consistently including correlations.
As a byproduct, we show that the cancellation of correlations between
the summed inputs to pairs of neurons does not originate from the
fast tracking of external input, but from the suppression of fluctuations
on the population level by the local network. The suppression of fluctuations
on the population level is a necessary constraint, but not sufficient
to determine the structure of correlations. Therefore, the structure
of correlations does not follow from the fast tracking of external
inputs.
\end{abstract}

\section*{Author summary}

The co-occurrence of action potentials of pairs of neurons within
short time intervals is known since long. Such synchronous events
can appear time-locked to the behavior of an animal and also theoretical
considerations argue for a functional role of synchrony. Early theoretical
work tried to explain correlated activity by neurons transmitting
common fluctuations due to shared inputs. This, however, overestimates
correlations. Recently the recurrent connectivity of cortical networks
was shown responsible for the observed low baseline correlations.
Two different explanations were given: One argues that excitatory
and inhibitory population activities closely follow the external inputs
to the network, so that their effects on a pair of cells mutually
cancel. Another explanation relies on negative recurrent feedback
to suppress fluctuations in the population activity, equivalent to
small correlations. In a biological neuronal network one expects both,
external inputs and recurrence, to affect correlated activity. The
present work extends the theoretical framework of correlations to
include both contributions and explains their qualitative differences.
Moreover the study shows that the arguments of fast tracking and recurrent
feedback are not equivalent, only the latter correctly predicts the
cell-type specific correlations.

\section{Introduction}

The spatio-temporal structure and magnitude of correlations in cortical
neural activity have since long been subject of research for a variety
of reasons: The experimentally observed task-dependent modulation
of correlations points at a potential functional role. In the motor
cortex of behaving monkeys, for example, synchronous action potentials
appear at behaviorally relevant time points \cite{Kilavik09_12653}.
The degree of synchrony is modulated by task performance, and the
precise timing of synchronous events follows a change of the behavioral
protocol after a phase of re-learning. In primary visual cortex, saccades
(eye movements) are followed by brief periods of synchronized neural
firing \cite{Maldonado08_1523,Ito11_2482}. Further, correlations
and fluctuations depend on the attentive state of the animal \cite{Crochet11_1160},
with higher correlations and slow fluctuations observed during quiet
wakefulness, and faster, uncorrelated fluctuations in the active state
\cite{Poulet08_881}. It is still unclear whether the observed modulation
of correlations is in fact employed by the brain, or whether it is
merely an epiphenomenon. Theoretical studies have suggested a number
of interpretations and mechanisms of how correlated firing could be
exploited: Correlations in afferent spike-train ensembles may provide
a gating mechanism by modulating the gain of postsynaptic cells (for
a review, see \cite{Salinas01}). Synchrony in afferent spikes (or,
more generally, synchrony in spike arrival) can enhance the reliability
of postsynaptic responses and, hence, may serve as a mechanism for
a reliable activation and propagation of precise spatio-temporal spike
patterns \cite{Abeles82,Diesmann99_529,Izhikevich06_245,Sterne12_2053}.
Further, it has been argued that synchronous firing could be employed
to combine elementary representations into larger percepts \cite{Hebb49,Malsburg81,Abeles82,Bienenstock95,Singer95}.
While correlated firing may constitute the substrate for some en-
and decoding schemes, it can be highly disadvantageous for others:
The number of response patterns which can be triggered by a given
afferent spike-train ensemble becomes maximal if these spike trains
are uncorrelated \cite{Tripp07_1830}. In addition, correlations in
the ensemble impair the ability of readout neurons to decode information
reliably in the presence of noise (see e.g. \cite{Zohary94_140,Tripp07_1830,Tetzlaff12_e1002596}).
Recent studies have indeed shown that biological neural networks implement
a number of mechanisms which can efficiently decorrelate neural activity,
such as the nonlinearity of spike generation \cite{DeLaRocha07_802},
synaptic-transmission variability and failure \cite{Rosenbaum11_1261,Rosenbaum13_475},
short-term synaptic depression \cite{Rosenbaum13_475}, heterogeneity
in network connectivity \cite{Bernacchia13_1732} and neuron properties
\cite{Padmanabhan10_1276}and the recurrent network dynamics \cite{Hertz10_427,Renart10_587,Tetzlaff12_e1002596}.
To study the significance of experimentally observed task-dependent
correlations, it is essential to provide adequate null hypotheses:
Which level and structure of correlations is to be expected in the
absence of any task-related stimulus or behavior? Even in the simplest
network models without time varying input, correlations in the neural
activity emerge as a consequence of shared input \cite{Shadlen98,Tetzlaff08_2133,Kriener08_2185}
and recurrent connectivity \cite{Renart10_587,Pernice11_e1002059,Tetzlaff12_e1002596,Trousdale12_e1002408,Helias13_023002}.
Irrespective of the functional aspect, the spatio-temporal structure
and magnitude of correlations between spike trains or membrane potentials
carry valuable information about the properties of the underlying
network generating these signals \cite{Tetzlaff08_2133,Pernice11_e1002059,Pernice12_031916,Trousdale12_e1002408,Helias13_023002}
and could therefore help constraining models of cortical networks.
Further, the quantification of spike-train correlations is a prerequisite
to understand how correlation sensitive synaptic plasticity rules,
such as spike-timing dependent plasticity \cite{Bi98}, interact with
the recurrent network dynamics \cite{Gilson09_1}. Finally, knowledge
of the expected level of correlations between synaptic inputs is crucial
for the correct interpretation of extracellular signals like the local-field
potential (LFP) \cite{Linden11_859}.

Previous theoretical studies on correlations in local cortical networks
provide analytical expressions for the magnitude \cite{Kriener08_2185,Renart10_587,Tetzlaff12_e1002596}
and the temporal shape \cite{Ginzburg94,Meyer02,Trousdale12_e1002408,Helias13_023002}
of average pairwise correlations, capture the influence of the connectivity
on correlations \cite{Lindner05_061919,Ostojic09_10234,Pernice11_e1002059,Pernice12_031916,Trousdale12_e1002408,Hu13_P03012},
and connect oscillatory network states emerging from delayed negative
feedback \cite{Brunel99} to the shape of correlation functions \cite{Helias13_023002}.
We have in particular shown recently that negative feedback loops,
abundant in cortical networks, constitute an efficient decorrelation
mechanism and therefore allow neurons to fire nearly independently
despite substantial shared presynaptic input \cite{Tetzlaff12_e1002596}
(see also \cite{Lindner05_061919,Renart10_587,LitvinKumar12_e1002667}).
We further pointed out that in networks of excitatory (E) and inhibitory
(I) neurons, the correlations between neurons of different cell type
(EE, EI, II) differ in both magnitude and temporal shape, even if
excitatory and inhibitory neurons have identical properties and input
statistics \cite{Tetzlaff12_e1002596,Helias13_023002}. It remains
unclear, however, how this cell-type specificity of correlations is
affected by the connectivity of the network.

The majority of previous theoretical studies on cortical circuits
is restricted to local networks driven by external sources representing
thalamo-cortical or cortico-cortical inputs (e.g. \cite{Vreeswijk96,Amit97,Brunel00_183}).
Most of these studies emphasize the role of the local network connectivity
(e.g. \cite{Potjans12_358}). Despite the fact that inputs from remote
(external) areas constitute a substantial fraction of all excitatory
inputs (about $50\%$ \cite{Abeles82}, see also \cite{Binzegger04,Stepanyants09_3555}),
their spatio-temporal structure is often abstracted by assuming that
neurons in the local network are independently driven by external
sources. A priori, this assumption can hardly be justified: neurons
belonging to the local cortical network receive, at least to some
extent, inputs from identical or overlapping remote areas, for example
due to patchy (clustered) horizontal connectivity \cite{Gilbert83,Voges10_277}.
Hence, shared-input correlations are likely to play a role not only
for local but also for external inputs. Coherent activation of neurons
in remote presynaptic areas constitutes another source of correlated
external input, in particular for sensory areas \cite{Okun_535_08,Poulet08_881,Gentet10_422,Crochet11_1160}.
So far, it is largely unknown how correlated external input affects
the dynamics of local cortical networks and alters correlations in
their neural activity.

In this article, we investigate how the magnitude and the cell-type
specificity of correlations depend on i) the connectivity in local
cortical networks of finite size and ii) the level of correlations
in external inputs. Existing theories of correlations in cortical
networks are not sufficient to address these questions as they either
do not incorporate correlated external input \cite{Ginzburg94,Tetzlaff12_e1002596,Trousdale12_e1002408,Pernice11_e1002059,Pernice12_031916}
or assume infinitely large networks \cite{Renart10_587}. Lindner
et al. \cite{Lindner05_061919} studied the responses of finite populations
of spiking neurons receiving correlated external input, but described
inhibitory feedback by a global compound process.

Our work builds on the existing theory of correlations in stochastic
binary networks \cite{Ginzburg94}, a well-established model in the
neuroscientific community \cite{Vreeswijk96,Renart10_587}. This model
has the advantage of requiring for its analytical treatment elementary
mathematical methods only. We employ the same network structure used
in the work by Renart et al. \cite{Renart10_587} which relates the
mechanism of recurrent decorrelation to the fast tracking of external
signals (see \cite{Parga13_P03010} for a recent review). This choice
enables us to reconsider the explanation of decorrelation by negative
feedback \cite{Tetzlaff12_e1002596}, originally shown for networks
of leaky integrate-and-fire neurons, and to compare it to the findings
of Renart et al. In fact, the motivation for the choice of the model
arose from the review process of \cite{Tetzlaff12_e1002596}, during
which both the reviewers and the editors encouraged us to elucidate
the relation of our work to the one of Renart et al. in a separate
subsequent manuscript. The present work delivers this comparison.

We show here that the results presented in \cite{Tetzlaff12_e1002596}
for the leaky integrate-and-fire model are in qualitative agreement
with those in networks of binary neurons. The formal relationship
between spiking models and the binary neuron model is established
in \cite{Grytskyy13_arxiv1304}. In particular, for weak correlations
it can be shown that both models map to the Ornstein-Uhlenbeck process
with one important difference: The location of the effective white
noise for spiking neurons is additive in the output, while for binary
neurons the effective noise is low-pass filtered, or equivalently
additive on the input side of the neuron.

The remainder of the manuscript is organized as follows: In \nameref{sec:Methods},
we develop the theory of correlations in recurrent random networks
of excitatory and inhibitory cells driven by fluctuating input from
an external population of finite size. We account for the fluctuations
in the synaptic input to each cell, which effectively linearize the
hard threshold of the neurons \cite{VanVreeswijk98_1321,Renart10_587}.
We further include the resulting finite-size correlations into the
established mean-field description \cite{Vreeswijk96,VanVreeswijk98_1321}
to increase the accuracy of the theory. In \nameref{sec:Results},
we first show in \nameref{sub:intri_extri} that correlations in recurrent
networks are not only caused by the externally imposed correlated
input, but also by intrinsically generated fluctuations of the local
populations. We demonstrate that the external drive causes an overall
shift of the correlations, but that their relative magnitude is mainly
determined by the intrinsically generated fluctuations. In \nameref{sub:cancellation_input},
we revisit the earlier reported phenomenon of the suppression of correlations
between input currents to pairs of cells \cite{Renart10_587} and
show that it is a direct consequence of the suppression of fluctuations
on the population level \cite{Tetzlaff12_e1002596}. In \nameref{sub:Limit-of-infinite}
we consider the strong coupling limit of the theory, where the network
size goes to infinity to recover earlier results for inhomogeneous
connectivity \cite{Renart10_587} and to extend these results to homogeneous
connectivity. Subsequently, in \nameref{sub:conn_struct_corr_struct},
we investigate in how far the reported structure of correlations is
a generic feature of balanced networks and isolate parameters of the
connectivity determining this structure. Finally, in \nameref{sec:Discussion},
we summarize our results and their implications for the interpretation
of experimental data, discuss the limitations of the theory, and provide
an outlook of how the improved theory may serve as a further building
block to understand processing of correlated activity.

\section{Methods\label{sec:Methods}}

\subsection{Networks of binary neurons\label{sub:Networks-of-binary}}

We denote the activity of neuron $i$ as $n_{i}(t)$. The state $n_{i}(t)$
of a binary neuron is either $0$ or $1$, where $1$ indicates activity,
$0$ inactivity \cite{Ginzburg94,Buice09_377,Renart10_587}. The state
of the network of $N$ such neurons is described by a binary vector
$\n=(n_{1},\ldots,n_{N})\in\{0,1\}^{N}$. We denote the mean activity
as $m_{i}=\langle n_{i}(t)\rangle_{t}$, the (zero time lag) covariance
of the activities of a pair $(i,j)$ of neurons is defined as $c_{ij}=\langle\delta n_{i}(t)\delta n_{j}(t)\rangle_{t}$,
where $\delta n_{i}(t)=n_{i}(t)-m_{i}$ is the deviation of neuron
$i$'s activity from expectation and the average $\langle\rangle_{t}$
is over time and realizations of the stochastic activity.

\begin{figure}
\centering{}\includegraphics[scale=0.5]{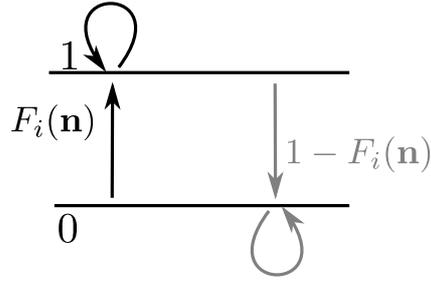}\caption{State transitions of a binary neuron. Each neuron is updated at random
time points, intervals are i.i.d. exponential with mean duration $\tau$,
so the rate of updates per neuron $i$ is $\tau^{-1}$. The probability
of neuron $i$ to end in the up-state ($1$) is determined by the
gain function $F_{i}(\n)$ which potentially depends on the states
$\n$ of all neurons in the network. The up-transitions are indicated
by black arrows. The probability for the down state ($0$) is given
by the complementary probability $1-F_{i}(\n)$, indicated by gray
arrows. \label{fig:binary_transition}}
\end{figure}
The neuron model shows stochastic transitions (at random points in
time) between the two states $0$ and $1$ controlled by transition
probabilities, as illustrated in \prettyref{fig:binary_transition}.
Using asynchronous update \cite{PDP86a}, in each infinitesimal interval
$[t,t+\delta t)$ each neuron in the network has the probability $\frac{1}{\tau}\delta t$
to be chosen for update \cite{Hopfield82}, where $\tau$ is the time
constant of the neuronal dynamics. An equivalent implementation draws
the time points of update independently for all neurons. For a particular
neuron, the sequence of update points has exponentially distributed
intervals with mean duration $\tau$, i.e. update times form a Poisson
process with rate $\tau^{-1}$. We employ the latter implementation
in the globally time-driven \cite{Hanuschkin10_113} spiking simulator
NEST \cite{Gewaltig_07_11204}, and use a discrete time resolution
$\delta t=0.1\ms$ for the intervals.  The stochastic update constitutes
a source of noise in the system. Given the $i$-th neuron is selected
for update, the probability to end in the up-state ($n_{i}=1$) is
determined by the gain function $F_{i}(\n)$ which possibly depends
on the activity $\n$ of all other neurons. The probability to end
in the down state ($n_{i}=0$) is $1-F_{i}(\n)$. This model has been
considered earlier \cite{Hertz91,Ginzburg94,Buice09_377}, and here
we follow the notation introduced in the latter work.

The stochastic system is completely characterized by the joint probability
distribution $p(\n)$ in all $N$ binary variables $\n$. Knowing
the joint probability distribution, arbitrary moments can be calculated,
among them pairwise correlations. Here we are only concerned with
the stationary state of the network. A stationary solution of $p(\n)$
implies that for each state a balance condition holds, so that the
incoming and outgoing probability fluxes sum up to zero. The occupation
probability of the state is then constant. We denote as $\n_{i+}=(n_{1},\ldots,n_{i-1},1,n_{i+1},\ldots,n_{N})$
the state, where the $i$-th neuron is active ($n_{i}=1$), and $\n_{i-}$
where neuron $i$ is inactive ($n_{i}=0$). Since in each infinitesimal
time interval at most one neuron can change state, for each given
state $\n$ there are $N$ possible transitions (each corresponding
to one of the $N$ neurons changing state). The sum of the probability
fluxes into the state and out of the state must compensate to zero
\cite{Kelly79}, so

\begin{eqnarray}
0=\tau\frac{\partial p(\n)}{\partial t} & = & \sum_{i=1}^{N}\underbrace{(2n_{i}-1)}_{\text{direction of flux}}\left(\underbrace{p(\n_{i-})F_{i}(\n_{i-})}_{\text{neuron }i\text{ transition up}}-\underbrace{p(\n_{i+})(1-F_{i}(\n_{i+}))}_{\text{neuron }i\text{ transition down}}\right)\quad\forall\quad\n\in\{0,1\}^{N}.\label{eq:balancing_flux}
\end{eqnarray}
From this equation we derive expressions for the first $\langle n_{k}\rangle$
and second moments $\langle n_{k}n_{l}\rangle$ by multiplying with
$n_{k}n_{l}$ and summing over all possible states $\n\in\{0,1\}^{N}$,
which leads to
\begin{eqnarray*}
0 & = & \sum_{\n\in\{0,1\}^{N}}\sum_{i=1}^{N}n_{k}n_{l}(2n_{i}-1)\underbrace{\left(p(\n_{i-})F_{i}(\n_{i-})-p(\n_{i+})(1-F_{i}(\n_{i+}))\right)}_{\equiv G_{i}(\n\backslash n_{i})}.
\end{eqnarray*}
Note that the term denoted $G_{i}(\n\backslash n_{i})$ does not depend
on the state of neuron $i$. We use the notation $\n\backslash n_{i}$
for the state of the network excluding neuron $i$, i.e. $\n\backslash n_{i}=(n_{1},\ldots,n_{i-1},n_{i+1},\ldots,n_{N})$.
Separating the terms in the sum over $i$ into those with $i\neq k,l$
and the two terms with $i=k$ and $i=l$, we obtain

\begin{eqnarray*}
0 & = & \sum_{\n}\;\sum_{i=1,i\neq k,l}^{N}n_{k}n_{l}(2n_{i}-1)\, G_{i}(\n\backslash n_{i})+n_{k}n_{l}(2n_{k}-1)\, G_{k}(\n\backslash n_{k})+n_{k}n_{l}(2n_{l}-1)\, G_{l}(\n\backslash n_{l})\\
 & = & \sum_{i=1,i\neq k,l}^{N}\;\sum_{\n\backslash n_{i}}n_{k}n_{l}(G_{i}(\n\backslash n_{i})-G_{i}(\n\backslash n_{i}))+\sum_{\n}n_{k}n_{l}\, G_{k}(\n\backslash n_{k})+\sum_{\n}n_{k}n_{l}\, G_{l}(\n\backslash n_{l})
\end{eqnarray*}
where we obtained the first term by explicitly summing over state
$n_{i}\in\{0,1\}$ (i.e. using $\sum_{\n\in\{0,1\}^{N}}=\sum_{\n\backslash n_{i}\in\{0,1\}^{N-1}}\sum_{n_{i}=0}^{1}$
and evaluating the sum $\sum_{n_{1}=0}^{1}$). This first sum obviously
vanishes. The remaining terms are of identical form with the roles
of $k$ and $l$ interchanged. We hence only consider the first of
them and obtain the other by symmetry. The first term simplifies to
\begin{eqnarray*}
 &  & \sum_{\n}n_{k}n_{l}\, G_{k}(\n\backslash n_{k})\stackrel{n_{k}=1}{=}\sum_{\n\backslash n_{k}}n_{l}\, G_{k}(\n\backslash n_{k})\\
 & \stackrel{\text{def. }G_{k}}{=} & \begin{cases}
\sum_{\n\backslash n_{k}}p(\n_{k-})\, F_{k}(\n_{k-})+p(\n_{k+})\, F_{k}(n_{k+})-p(\n_{k+}) & \quad\text{for }k=l\\
\sum_{\n\backslash n_{k}}p(\n_{k-})\, n_{l}\, F_{k}(\n_{k-})+p(\n_{k+})\, n_{l}\, F_{k}(n_{k+})-n_{l}\, p(\n_{k+}) & \quad\text{for }k\neq l
\end{cases}\\
 & = & \begin{cases}
\langle F_{k}(\n)\rangle-\langle n_{k}\rangle & \text{for }k=l\\
\langle F_{k}(\n)\, n_{l}\rangle-\langle n_{k}n_{l}\rangle & \text{for }k\neq l
\end{cases},
\end{eqnarray*}
where we denote as $\langle f(\n)\rangle=\sum_{\n\in\{0,1\}^{N}}p(\n)f(\n)$
the average of a function $f(\n)$ with respect to the distribution
$p(\n)$. Taken together with the mirror term $k\leftrightarrow l$,
we arrive at two conditions, one for the first ($k=l$, $\langle n_{k}^{2}\rangle=\langle n_{k}\rangle$)
and one for the second ($k\neq l$) moment
\begin{eqnarray}
2\langle n_{k}n_{l}\rangle & = & \begin{cases}
2\langle F_{k}(\n)\rangle & \text{for }k=l\\
\langle F_{k}(\n)\, n_{l}\rangle+\langle F_{l}(\n)\, n_{k}\rangle & \text{for }k\neq l
\end{cases}.\label{eq:mean_activity}
\end{eqnarray}
Considering the covariance $c_{kl}=\langle\delta n_{k}\delta n_{l}\rangle$
with centralized variables $\delta n_{k}=n_{k}-\langle n_{k}\rangle$,
for $k\neq l$ one arrives at
\begin{eqnarray}
2c_{kl} & = & \langle F_{k}(\n)\,\delta n_{l}\rangle+\langle F_{l}(\n)\,\delta n_{k}\rangle.\label{eq:correlation_non_lin}
\end{eqnarray}
This equation is identical to eq.\ 3.9 in \cite{Ginzburg94}, to
eqs.\ 3.12 and 3.13 in \cite{Buice09_377}, and to eqs.\ (19)-(22)
in \cite[Supplementary material]{Renart10_587}.

\subsection{Mean-field solution\label{sub:Mean-field-solution}}

Starting from \eqref{eq:balancing_flux} for the general case $\frac{\partial p(\n,t)}{\partial t}\neq0$,
a similar calculation as the one resulting in \eqref{eq:mean_activity}
for $k=l$ leads to 
\begin{eqnarray*}
\tau\frac{\partial}{\partial t}\langle n_{k}\rangle & = & \langle F_{k}(\n)\rangle-\langle n_{k}\rangle,
\end{eqnarray*}
where we used $\langle n_{k}^{2}\rangle=\langle n_{k}\rangle$, valid
for binary variables. As in \cite{Renart10_587} we now assume a particular
form for the gain function and for the coupling between neurons by
specifying
\begin{eqnarray*}
F_{k}(\n) & = & H\left(h_{k}-\theta\right)\\
h_{k} & = & \sum_{l=1}^{N}J_{kl}n_{l}\\
H(x) & = & \begin{cases}
1 & \text{if }x\ge0\\
0 & \text{if }x<0
\end{cases},
\end{eqnarray*}
where $J_{kl}$ is the incoming synaptic weight from neuron $l$ to
neuron $k$, $H$ is the Heaviside function, and $\theta$ is the
threshold of the activation function. For positive $\theta$ the neuron
gets activated only if sufficient excitatory input is present and
for negative $\theta$ the neuron is intrinsically active even in
the absence of excitatory input. We denote by $h_{k}$ the summed
synaptic input to the neuron, sometimes also called the ``field''.
Because $n_{k}^{2}=n_{k}$, the variance $a_{k}$ of a binary variable
is $a_{k}\equiv\langle n_{k}^{2}\rangle-\langle n_{k}\rangle^{2}=(1-\langle n_{k}\rangle)\langle n_{k}\rangle.$
We now aim to solve \eqref{eq:mean_activity} for the case $k=l$,
i.e. the equation $\langle n_{k}\rangle=\langle F_{k}\rangle$. In
general, the right hand side depends on the fluctuations of all neurons
projecting to neuron $k$. An exact solution is therefore complicated.
However, for sufficiently irregular activity in the network we assume
the neurons to be approximately independent. Further assume that in
a network of homogeneous populations $\alpha$ (same parameters $\tau$,
$\theta$ and same statistics of the incoming connections for all
neurons, i.e. same number $K_{\alpha\beta}$ and strength $J_{\alpha\beta}$
of incoming connections from neurons in a given population $\beta$)
the mean activity of an individual neuron can be represented by the
population mean $m_{\alpha}=\langle\frac{1}{N_{\alpha}}\sum_{i\in\alpha}n_{i}\rangle$.
The mean input to a neuron in population $\alpha$ then is 
\begin{eqnarray}
\langle h_{\alpha}\rangle & = & \sum_{\beta}K_{\alpha\beta}J_{\alpha\beta}m_{\beta}\equiv\mu_{\alpha}.\label{eq:mu_alpha}
\end{eqnarray}
We assumed in the last step identical synaptic amplitudes $J_{\alpha\beta}$
for a synapse from a neuron in population $\beta$ to a neuron in
population $\alpha$. So the input to each neuron has the same mean
$\langle h_{\alpha}\rangle$. As a first approximation, if the mean
activity in the network is not saturated, i.e. neither $0$ nor $1$,
mapping this activity back by the inverse gain function to the input,
$h_{\alpha}$ must be close to the threshold value, so
\begin{eqnarray}
\langle h_{\alpha}\rangle & \simeq & \theta.\label{eq:cancellation_mu}
\end{eqnarray}
This relation may be solved for $m_{E}$ and $m_{I}$ to obtain a
coarse estimate of the activity in the network \cite{Vreeswijk96,VanVreeswijk98_1321}.
In mean-field approximation we assume that the fluctuations of the
fields of individual neurons $h_{\alpha}$ around their mean are mutually
independent, so that the fluctuations $\delta h_{\alpha}=h_{\alpha}-\langle h_{\alpha}\rangle$
of $h_{\alpha}$ are, in turn, caused by a sum of independent random
variables and hence the variances add up to the variance $\sigma_{\alpha}^{2}$
of the field 
\begin{eqnarray}
\langle\delta h_{\alpha}^{2}\rangle & = & \sum_{\beta}K_{\alpha\beta}J_{\alpha\beta}^{2}m_{\beta}(1-m_{\beta})\equiv\sigma_{\alpha}^{2}.\label{eq:sigma_alpha}
\end{eqnarray}
As $h_{\alpha}$ is a sum of typically thousands of synaptic inputs,
it approaches a Gaussian distribution $h_{\alpha}\sim\mathcal{N}(\mu_{\alpha},\sigma_{\alpha}^{2})$
with mean $\mu_{\alpha}$ and variance $\sigma_{\alpha}^{2}$. In
this approximation the mean activity in the network is the solution
of 
\begin{eqnarray}
\tau\frac{\partial}{\partial t}m_{\alpha}+m_{\alpha} & = & \langle F_{\alpha}(m_{E},m_{I},m_{x})\rangle\quad\forall\;\alpha\in\{E,I\}\nonumber \\
 & \simeq & \int_{-\infty}^{\infty}H(x-\theta)\:\mathcal{N}(\mu_{\alpha},\sigma_{\alpha}^{2},x)\, dx\nonumber \\
 & = & \int_{\theta}^{\infty}\mathcal{N}(\mu_{\alpha},\sigma_{\alpha}^{2},x)\, dx\nonumber \\
 & = & \frac{1}{2}\erfc\left(\frac{\theta-\mu_{\alpha}}{\sqrt{2}\sigma_{\alpha}}\right).\label{eq:mean_field_activity}
\end{eqnarray}
This equation needs to be self-consistently solved with $\frac{\partial m_{\alpha}}{\partial t}=0$
by numerical or graphical methods in order to obtain the stationary
activity, because $\mu_{\alpha}(m_{E},m_{I},m_{x})$ and $\sigma_{\alpha}(m_{E},m_{I},m_{x})$
depend on $m_{\alpha}\forall\alpha\in\{E,I,X\}$ themselves. We here
employ the algorithm $\mathrm{hybrd}$ and $\mathrm{hybrj}$ from
the MINPACK package, implemented in scipy (version 0.9.0) \cite{scipy01}
as the function $\mathrm{scipy.optimize.fsolve}$.

\subsection{Linearized equation for correlations and susceptibility}

In general, the term $\langle F_{k}(\n)\,\delta n_{l}\rangle$ in
\eqref{eq:correlation_non_lin} couples moments of arbitrary order,
resulting in a moment hierarchy \cite{Buice09_377}. Here we only
determine an approximate solution. Since the single synaptic amplitudes
$J_{ki}$ are small, we linearize the effect of a single synaptic
input. We apply the linearization to the two terms of the form $\langle F_{k}(\n)\,\delta n_{l}\rangle$
on the right hand side of \eqref{eq:correlation_non_lin}. In the
recurrent network, the activity of each neuron in the vector $\n$
may be correlated to the activity of any other neuron $n_{i}$. Therefore,
the input $h_{k}$ sensed by neuron $k$ not only depends on $n_{l}$
directly, but also indirectly through the correlations of $n_{l}$
with any of the other neurons $n_{i}$ that project to neuron $k$.
We need to take this dependence into account in the linearization.
Considering the effect of one particular input $n_{i}$ explicitly
one gets

\begin{eqnarray*}
\langle F_{k}(\n)\delta n_{l}\rangle & = & \langle H(h_{k}-\theta)\,\delta n_{l}\rangle\\
 & = & \langle H(h_{k\backslash n_{i}}+J_{ki}-\theta)\, n_{i}\delta n_{l}+H(h_{k\backslash n_{i}}-\theta)\,(1-n_{i})\,\delta n_{l}\rangle\\
 & = & \langle(H(h_{k\backslash n_{i}}+J_{ki}-\theta)-H(h_{k\backslash n_{i}}-\theta))\, n_{i}\delta n_{l}\rangle+\langle H(h_{k\backslash n_{i}}-\theta)\,\delta n_{l}\rangle.
\end{eqnarray*}
The first term $\langle(H(h_{k\backslash n_{i}}+J_{ki}-\theta)-H(h_{k\backslash n_{i}}-\theta))\, n_{i}\delta n_{l}\rangle$
already contains two factors $n_{i}$ and $\delta n_{l}$, so it takes
into account second order moments. Performing the expansion for the
next input would yield terms corresponding to correlations of higher
order, which are neglected here. This amounts to the assumption that
the remaining fluctuations in $h_{k\backslash n_{i}}$ are independent
of $n_{i}$ and $n_{l}$, and we again approximate them by a Gaussian
random variable $x\sim\mathcal{N}(\mu_{k},\sigma_{k})$ with mean
$\mu_{k}=\langle h_{k}\rangle$ and variance $\sigma_{k}^{2}=\langle\delta h_{k}^{2}\rangle$,
so $\langle(H(x+J_{ki}-\theta)-H(x-\theta))\rangle_{x}\langle n_{i}\delta n_{l}\rangle_{\n}\simeq S(\mu_{k},\sigma_{k})\, J_{ki}\,\langle n_{i}\delta n_{l}\rangle_{\n}+O(J_{ki}^{2}).$
Here we used the smallness of the synaptic weight $J_{ki}$ and replaced
the difference by the derivative $S(\mu_{k},\sigma_{k})=\left.\frac{\partial\langle H(x+J)\rangle_{x\sim\mathcal{N}(\mu_{k},\sigma_{k})}}{\partial J}\right|_{J=0}$,
which has the form of a susceptibility. Using the explicit expression
for the Gaussian integral \eqref{eq:mean_field_activity}, the susceptibility
is exactly
\begin{eqnarray}
S(\mu_{k},\sigma_{k}) & = & \frac{1}{\sqrt{2\pi}\sigma_{k}}e^{-\frac{(\mu_{k}-\theta)^{2}}{2\sigma_{k}^{2}}}.\label{eq:susceptibility}
\end{eqnarray}
The same expansion holds for the remaining inputs to cell $k$. With
$\langle n_{i}\delta n_{l}\rangle=\begin{cases}
a_{i} & \text{for }i=l\\
c_{il} & \text{for }i\neq l
\end{cases}$, the equation for the pairwise correlations \eqref{eq:correlation_non_lin}
in linear approximation takes the form
\begin{eqnarray}
2c_{kl} & = & S(\mu_{k},\sigma_{k})\,\left(\sum_{j}J_{kj}c_{jl}+J_{kl}a_{l}\right)+S(\mu_{l},\sigma_{l})\,\left(\sum_{j}J_{lj}c_{jk}+J_{lk}a_{k}\right),\label{eq:corr_lin}
\end{eqnarray}
corresponding to eq.\ (6.8) in \cite{Ginzburg94} and eqs.\ (31)-(33)
in \cite[Supplementary material]{Renart10_587}. Note, however, that
the linearization used in \cite{Ginzburg94} relies on the smoothness
of the gain function due to additional local noise, whereas here and
in \cite[Supplementary material]{Renart10_587} a Heaviside gain function
is used and only the existence of noise generated by the network itself
justifies the linearization. If the input to each neuron is homogeneous,
i.e. $\mu_{k}=\mu_{\alpha}$ and $\sigma_{k}=\sigma_{\alpha}$ for
all neurons $k$ in population $\alpha$, a structurally similar equation
connects the correlations $c_{\alpha\beta}=\frac{1}{N_{\alpha}N_{\beta}}\sum_{k\in\alpha,l\in\beta,k\neq l}\, c_{kl}$
averaged over disjoint pairs of neurons belonging to two (possibly
identical) populations $\alpha$,$\beta$ with the population averaged
variances $a_{\alpha}=\frac{1}{N_{\alpha}}\sum_{k\in\alpha}a_{k}$
\begin{eqnarray}
2c_{\alpha\beta} & = & \sum_{\gamma\in\{E,I,X\}}\left(w_{\alpha\gamma}c_{\gamma\beta}+w_{\beta\gamma}c_{\gamma\alpha}\right)+w_{\alpha\beta}\frac{a_{\beta}}{N_{\beta}}+w_{\beta\alpha}\frac{a_{\alpha}}{N_{\alpha}}\label{eq:population_correlation}\\
\text{with }w_{\alpha\beta} & = & S(\mu_{\alpha},\sigma_{\alpha})\, J_{\alpha\beta}\, K_{\alpha\beta}.\nonumber 
\end{eqnarray}
In deriving the last expression, we replaced variances of individual
neurons and correlations between individual pairs by their respective
population averages and counted the number of connections. This equation
corresponds to eqs.\ (9.14)-(9.16) in \cite{Ginzburg94} (which lack,
however, the external population $X$, and note the typo in the first
term in line 2 of eq.\ (9.16), which should read $-\frac{1}{2}\bar{J}{}_{EI}C_{II}(0)$)
and eqs.\ (36) in \cite[Supplementary material]{Renart10_587}.
Written in matrix form \eqref{eq:population_correlation} takes the
form \eqref{eq:corrected_corr_structure} of the main text, where
we defined

\begin{eqnarray}
A & = & \left(\begin{array}{ccc}
2-2w_{EE} & -2w_{EI} & 0\\
-w_{IE} & 2-\left(w_{EE}+w_{II}\right) & -w_{EI}\\
0 & -2w_{IE} & 2-2w_{II}
\end{array}\right)\nonumber \\
B & = & \left(\begin{array}{cc}
2w_{EE} & 0\\
w_{IE} & w_{EI}\\
0 & 2w_{II}
\end{array}\right)\qquad C=\left(\begin{array}{cc}
2w_{EX} & 0\\
w_{IX} & w_{EX}\\
0 & 2w_{IX}
\end{array}\right)\nonumber \\
D & = & \left(\begin{array}{cc}
2-w_{EE} & -w_{EI}\\
-w_{IE} & 2-w_{II}
\end{array}\right)\qquad E=\left(\begin{array}{c}
w_{EX}\\
w_{IX}
\end{array}\right).\label{eq:matrix_form_pop_correlation}
\end{eqnarray}
The explicit solution of the latter system of \eqref{eq:corrected_corr_structure}
is
\begin{eqnarray}
\left(\begin{array}{c}
c_{XE}\\
c_{XI}
\end{array}\right) & = & \frac{1}{(2-w_{EE})(2-w_{II})-w_{EI}w_{IE}}\left(\begin{array}{c}
(2-w_{II})w_{EX}+w_{EI}w_{IX}\\
(2-w_{EE})w_{IX}+w_{IE}w_{EX}
\end{array}\right)\,\frac{a_{X}}{N_{X}}.\label{eq:external_correlation}
\end{eqnarray}

\subsection{Mean-field theory including finite-size correlations\label{sub:self_consistent_mf_corr}}

The mean-field solution presented in\\
\nameref{sub:Mean-field-solution} assumes that correlations among
the neurons in the network are negligible. This assumption enters
the expression \eqref{eq:sigma_alpha} for the variance of the input
to a neuron. Having determined the actual magnitude of the correlations
in \eqref{eq:corrected_corr_structure}, we are now able to state
a more accurate approximation in which we take these correlations
into account, modifying the expression for the variance of the field
$h_{\alpha}$
\begin{eqnarray}
\sigma_{\alpha}^{2} & = & \sum_{\beta\in\{E,I,X\}}K_{\alpha\beta}J_{\alpha\beta}^{2}m_{\beta}(1-m_{\beta})+\sum_{\beta,\gamma\in\{E,I,X\}}(KJ)_{\alpha\beta}(KJ)_{\alpha\gamma}c_{\beta\gamma}\label{eq:variance_corr}\\
\text{with }(KJ)_{\alpha\beta} & \equiv & K_{\alpha\beta}J_{\alpha\beta}.\nonumber 
\end{eqnarray}
This correction suggests an iterative scheme: Initially we solve the
mean-field equation \eqref{eq:mean_field_activity} assuming $c_{\alpha\beta}=0$
(hence $\sigma_{\alpha}$ given by \eqref{eq:sigma_alpha}). In each
step of the iteration we then calculate the correlations by \eqref{eq:corrected_corr_structure},
compute the mean-field solution of \eqref{eq:mean_field_activity}
and the susceptibility $S(\mu_{\alpha},\sigma_{\alpha})$ \eqref{eq:susceptibility},
taking into account the correlations \eqref{eq:variance_corr} determined
in the previous step. These steps are iterated until the solution
($m_{\alpha},c_{\alpha\beta}\quad\forall\alpha,\beta$) converges.
We use this approach to determine the correlation structure in \prettyref{fig:correlations_ext},
where we iterated until the solution became invariant up to a residual
absolute difference of $10^{-15}$. A comparison of the distribution
of the total synaptic input $h_{E}$ at the end of the iteration with
a Gaussian distribution with parameters $\mu_{E}$ and $\sigma_{E}$
is shown in \prettyref{fig:correlations_ext}D.

\subsection{Influence of inhomogeneity of in-degrees\label{sec:Influence-of-inhomogeneity}}

In the previous sections we assumed the number of incoming connections
to be the same for all neurons. Studying a random network in its original
Erdös-Rényi \cite{Palmer85} sense, the number of synaptic inputs
$K_{i\beta}$ to a neuron $i\in\alpha$ from population $\beta$ is
a binomially distributed random number. As a consequence, the time-averaged
activity differs among neurons. Since each neuron $i\in\alpha$ samples
a random subset of inputs from a given population $\beta$, we can
assume that the realization of $K_{i\beta}$ is independent of the
realization of the time-averaged activity of the inputs from population
$\beta$. So these two contributions to the variability of the mean
input $\delta\mu_{\alpha}^{2}$ add up. The number of incoming connections
to a neuron in population $i\in\alpha$ follows a binomial distribution
\begin{align*}
K_{i\beta} & \sim B(N_{\beta},p),
\end{align*}
where $p$ is the connection probability and $N_{\beta}$ the size
of the sending population. The mean value is as before $K_{\alpha\beta}=[\frac{1}{N_{\alpha}}\sum_{i\in\alpha}K_{i\beta}]=pN_{\beta}$,
where we denote the expectation value with respect to the realization
of the connectivity as $[]$. The variance of the in-degree is hence

\begin{align*}
\delta K_{\alpha\beta}^{2} & =\left[\frac{1}{N_{\alpha}}\sum_{i\in\alpha}\left(K_{i\beta}-K_{\alpha\beta}\right)^{2}\right]=N_{\beta}p(1-p)=K_{\alpha\beta}(1-p).
\end{align*}
In the following we adapt the results from \cite{VanVreeswijk98_1321,Renart10_587}
to the present notation. The contribution of the variability of the
number of synapses to the variance of the mean input is $\sum_{\beta}\, J_{\alpha\beta}^{2}\delta K_{\alpha\beta}^{2}m_{\beta}^{2}$.
The contribution from the distribution of the mean activities can
be expressed by the variance of the mean activity defined as
\begin{align*}
\delta m_{\alpha}^{2} & \equiv\left[\frac{1}{N_{\alpha}}\sum_{i\in\alpha}m_{i}^{2}\right]-m_{\alpha}^{2}\\
 & \equiv q_{\alpha}-m_{\alpha}^{2}.
\end{align*}
The $K_{\alpha\beta}$ independently drawn inputs hence contribute
$\sum_{\beta}J_{\alpha\beta}^{2}K_{\alpha\beta}\delta m_{\beta}^{2}$,
as the variances of the $K_{\alpha\beta}$ terms add up. So together
we have \cite[eq. 5.5 - 5.6]{VanVreeswijk98_1321}

\begin{align*}
\delta\mu_{\alpha}^{2} & =\sum_{\beta}\, J_{\alpha\beta}^{2}(\delta K_{\alpha\beta}^{2}m_{\beta}^{2}+K_{\alpha\beta}\delta m_{\beta}^{2}).
\end{align*}
Using $K_{\alpha\beta}=N_{\beta}p$ we obtain

\begin{align}
\delta\mu_{\alpha}^{2} & =\sum_{\beta}\, J_{\alpha\beta}^{2}\left(\delta K_{\alpha\beta}^{2}m_{\beta}{}^{2}+K_{\alpha\beta}\delta m_{\beta}^{2}\right)\nonumber \\
 & =\sum_{\beta}\, J_{\alpha\beta}^{2}K_{\alpha\beta}\left((1-p)\, m_{\beta}^{2}+q_{\beta}-m_{\beta}^{2}\right)\nonumber \\
 & =\sum_{\beta}\, J_{\alpha\beta}^{2}K_{\alpha\beta}\left(q_{\beta}-p\, m_{\beta}^{2}\right).\label{eq:variance_mean}
\end{align}
The latter expression differs from \cite[eq. 5.7]{VanVreeswijk98_1321}
only in the term $-pm_{\beta}^{2}$ that is absent in the work of
van Vreeswijk and Sompolinsky, because they assumed the number of
synapses to be Poisson distributed in the limit of sparse connectivity
\cite[Appendix, (A.6)]{VanVreeswijk98_1321} (also note that their
$J_{kl}$ corresponds to our $\sqrt{K_{\alpha\beta}}J_{\alpha\beta}$).
The expression \prettyref{eq:variance_mean} is identical to \cite[Supplementary, eq. (25)]{Renart10_587}.

Since the variance of a binary signal with time-averaged activity
$m_{i}$ is $m_{i}(1-m_{i})$, the population-averaged variance is
hence 
\begin{align}
a_{\alpha}=\frac{1}{N_{\alpha}}\sum_{i\in\alpha}[m_{i}(1-m_{i})] & =m_{\alpha}-q_{\alpha}.\label{eq:variance_binary_dist}
\end{align}
So the sum of $K_{\alpha\beta}$ such (uncorrelated) signals contributes
to the fluctuation of the input as

\begin{align}
\sigma_{\alpha}^{2}=[\delta h_{\alpha}^{2}] & =\sum_{\beta}\, J_{\alpha\beta}^{2}K_{\alpha\beta}(m_{\beta}-q_{\alpha}).\label{eq:variance_h}
\end{align}
The contribution due to the variability of the number of synapses
$\delta K_{\alpha\beta}^{2}$ can be neglected in the limit of large
networks \cite{Renart10_587}. The time-averaged activity of a single
cell with mean input $\mu_{i}$ and variance $\sigma_{i}^{2}$ is
given, as before by \eqref{eq:mean_field_activity} $m_{i}=\Phi(\mu_{i},\sigma_{i})$,
so the distribution of activity in the population is
\begin{align}
p(m) & =\int_{-\infty}^{\infty}\delta(m-\Phi(x,\sigma_{\alpha}))\,\mathcal{N}(\mu_{\alpha},\delta\mu_{\alpha}^{2},x)\, dx\nonumber \\
 & =\left(\Phi^{\prime}\right)^{-1}(\Phi^{-1}(m))\,\mathcal{N}(\mu_{\alpha},\delta\mu_{\alpha}^{2},\Phi^{-1}(m)).\label{eq:rate_distribution}
\end{align}
The mean activity of the whole population is
\begin{align}
m_{\alpha} & =\int_{-\infty}^{\infty}\mathcal{N}(\mu_{\alpha},\delta\mu_{\alpha}^{2},y)\,\Phi(y,\sigma_{\alpha}^{2})\, dy\nonumber \\
 & =\int_{-\infty}^{\infty}\mathcal{N}(\mu_{\alpha},\delta\mu_{\alpha}^{2},y)\,\int_{\theta}^{\infty}\mathcal{N}(y,\sigma_{\alpha}^{2},x)\, dx\, dy\nonumber \\
 & =\int_{\theta}^{\infty}\int_{-\infty}^{\infty}\mathcal{N}(\mu_{\alpha},\delta\mu_{\alpha}^{2},y)\,\mathcal{N}(y,\sigma_{\alpha}^{2},x)\, dy\, dx\nonumber \\
 & =\Phi(\mu_{\alpha},\sigma_{\alpha}^{2}+\delta\mu_{\alpha}^{2}),\label{eq:first_moment_rate_distribution}
\end{align}
because the penultimate line is a convolution of two Gaussian distributions,
so the means and variances add up. The second moment of the population
activity is
\begin{align}
q_{\alpha} & =\int_{-\infty}^{\infty}\mathcal{N}(\mu_{\alpha},\delta\mu_{\alpha}^{2},x)\,\Phi^{2}(x,\sigma_{\alpha}^{2})\, dx.\label{eq:second_moment_rate_distribution}
\end{align}
These expressions are identical to \cite[Supplementary, eqs. (26), (27)]{Renart10_587}.
The system of equations \eqref{eq:mu_alpha}, \eqref{eq:variance_mean},
\eqref{eq:variance_h}, \eqref{eq:first_moment_rate_distribution},
and \eqref{eq:second_moment_rate_distribution} can be solved self-consistently.
We use the algorithm $\mathrm{hybrd}$ and $\mathrm{hybrj}$ of the
MINPACK package, implemented in scipy (version 0.9.0) \cite{scipy01}
as the function $\mathrm{scipy.optimize.fsolve}$. This yields the
self-consistent solutions for $m_{\alpha}$ and $q_{\alpha}$ and
hence the distribution of time averaged activity \eqref{eq:rate_distribution}
can be obtained, shown in \prettyref{fig:renart_binary}F.

\section{Results\label{sec:Results}}

\begin{figure}
\centering{}\includegraphics[scale=2]{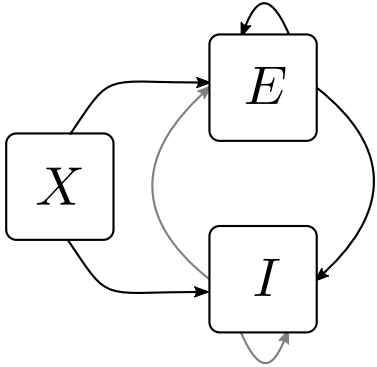}\caption{Recurrent local network of two populations of excitatory ($E$) and
inhibitory ($I$) neurons driven by a common external population ($X$).
The external population $X$ delivers stochastic activity to the local
network. The local network is a recurrent Erd\H{o}s-R\'{e}nyi random
network with homogeneous synaptic weights $J_{\alpha\beta}$ coupling
neurons in population $\beta$ to neurons in population $\alpha$,
for $\alpha,\beta\in\{E,I\}$ and same parameters for all neurons.
There are $N=8192$ neurons in both the excitatory and the inhibitory
population. The connection probability is $p=0.2$, and each neuron
in population $\alpha$ receives the same number $K=pN$ of excitatory
and inhibitory synapses. The size $N_{X}$ of the external population
determines the amount of shared input received by each pair of cells
in the local network. The neurons are modeled as binary units with
a hard threshold $\theta$.\label{fig:E_I_X_network}}
\end{figure}

Our aim is to investigate the effect of recurrence and external input
on the magnitude and structure of cross-correlations between the activities
in a recurrent random network, as defined in \nameref{sub:Networks-of-binary}.
We employ the established recurrent neuronal network model of binary
neurons in the balanced regime \cite{Vreeswijk96}. The binary dynamics
has the advantage to be more easily amendable to analytical treatment
than spiking dynamics. A method to calculate the pairwise correlations
exists since long \cite{Ginzburg94}. The choice of binary dynamics
moreover renders our results directly comparable to the recent findings
on decorrelation in such networks \cite{Renart10_587}. Our model
consists of three populations of neurons, one excitatory and one inhibitory
population which together represent the local network, and an external
population providing additional excitatory drive to the local network,
as illustrated in \prettyref{fig:E_I_X_network}. The external population
may either be conceived as representing input into the local circuit
from remote areas or as representing sensory input. The external population
contains $N_{X}$ neurons, which are pairwise uncorrelated and have
a stochastic activity with mean $m_{X}$. Each neuron in population
$\alpha\in\{E,I\}$ within the local network draws $K=pN$ connections
randomly from the finite pool of $N_{X}$ external neurons. $N_{X}$
therefore determines the number of shared afferents received by each
pair of cells from the external population with on average $K^{2}/N_{X}$
common synapses. In the extreme cases $N_{X}=K$ all neurons receive
exactly the same input, whereas for large $N_{X}\rightarrow\infty$
the fraction of shared external input approaches $0$. The common
fluctuating input received from the finite-sized external population
hence provides a signal imposing pairwise correlations, the amount
of which is controlled by the parameter $N_{X}$.

\subsection{Correlations are driven by intrinsic and external fluctuations\label{sub:intri_extri}}

\begin{figure}
\noindent \centering{}\includegraphics[scale=0.9]{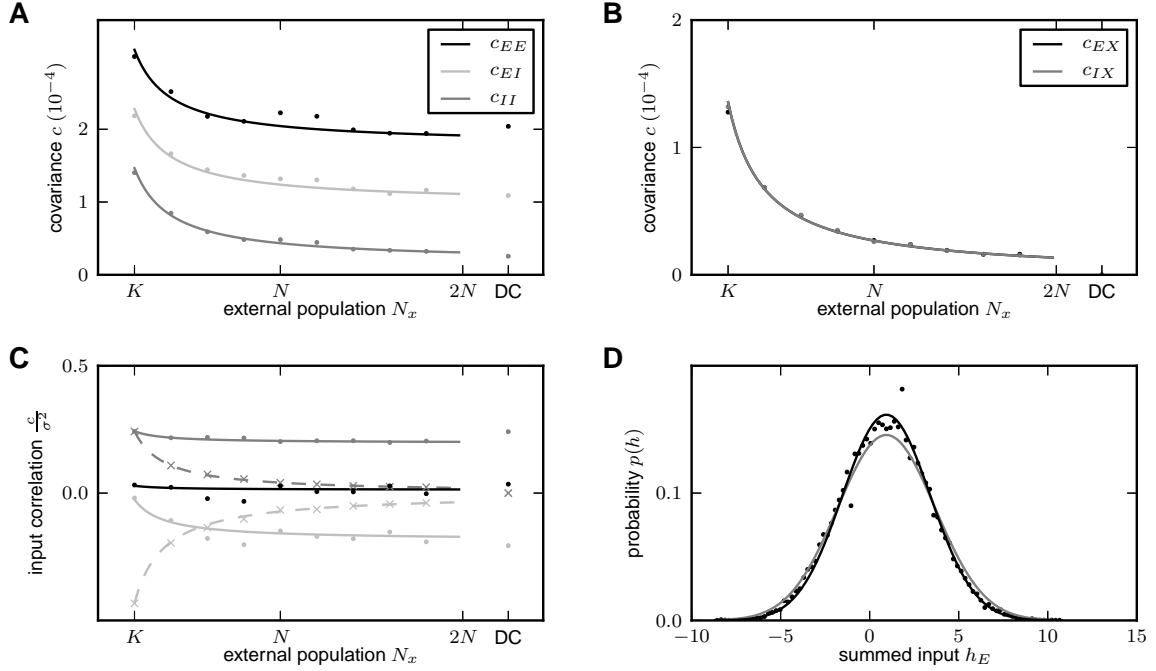}\caption{Correlations in a network of three populations as illustrated in \prettyref{fig:E_I_X_network}
in dependence of the size $N_{x}$ of the external population. Each
neuron in population $\alpha\in\{E,I\}$ receives $pN$ randomly drawn
excitatory inputs with weight $J_{\alpha E}=\frac{5}{\sqrt{N}}$,
$pN$ randomly drawn inhibitory inputs of weight $J_{\alpha I}=-\frac{10}{\sqrt{N}}$
and $pN$ external inputs of weight $J_{\alpha X}=\frac{5}{\sqrt{N}}$
(homogeneous random network with fixed in-degree, connection probability
$p=0.2$). \textbf{A} Correlations averaged over pairs of neurons
within the local network \eqref{eq:pop_avg_crosscorr}. Dots indicate
results of direct simulation over $T=30\second$ averaged over $(N/2)^{2}$
pairs of neurons. Curves show the analytical result \eqref{eq:corrected_corr_structure}.
The point ``DC'' shows the correlation structure emerging if the
drive from the external population is replaced by a constant value
$KJ_{\alpha X}m_{X}$, which provides the same mean input as the original
external drive. \textbf{B} Correlations between neurons within the
local network and the external population averaged over pairs of neurons
(same labeling as in A). \textbf{C} Correlation between the inputs
to a pair of cells in the network decomposed into the contributions
due to shared inputs $c_{\mathrm{shared}}$ (gray, eq.\ \ref{eq:c_shared})
and due to correlations $c_{\mathrm{corr}}$ in the presynaptic activity
(light gray, eq.\ \ref{eq:c_corr}). Dashed curves and St. Andrew's
Crosses show the contribution due to external inputs, solid curves
and dots show the contribution from local inputs. The sum of all components
is shown by black dots and curve. Curves are theoretical results based
on \eqref{eq:corrected_corr_structure}, \eqref{eq:c_shared}, and
\eqref{eq:c_corr}, symbols are obtained from simulation. \textbf{D}
Probability distribution of the fluctuating input $h_{E}$ to a single
neuron in the excitatory population. Dots show the histogram obtained
from simulation binned over the interval $[\min(h_{E}),\max(h_{E})]$
with a bin size of $-2J_{\alpha I}$. The gray curve is the prediction
of a Gaussian distribution obtained from mean-field theory neglecting
correlations, with mean and variance given by \eqref{eq:mu_alpha}
and \eqref{eq:sigma_alpha}, respectively. The black curve takes correlations
in the afferent signals into account and has a variance given by \eqref{eq:variance_corr}.
Other parameters: simulation resolution $\Delta t=0.1\ms$, synaptic
delay $d=\Delta t$, activity measurement in intervals of $1\ms$.
Threshold of the neurons $\theta=1$, time constant of inter-update
intervals $\tau=10\ms$. The average activity in the network is $m_{E}\simeq m_{I}\simeq m_{X}=0.5$.\label{fig:correlations_ext}}
\end{figure}

To explain the correlation structure observed in a network with external
inputs (\prettyref{fig:E_I_X_network}), we extend the existing theory
of pairwise correlations \cite{Ginzburg94} to include the effect
of externally imposed correlations. The global behavior of the network
can be studied with the help of the mean-field equation \eqref{eq:mean_field_activity}
for the population-averaged mean activity $m_{\alpha}=N_{\alpha}^{-1}\sum_{i\in\alpha}\langle n_{i}\rangle$
\begin{eqnarray}
m_{\alpha} & = & \frac{1}{2}\erfc\left(\frac{\theta-\mu_{\alpha}}{\sqrt{2}\sigma_{\alpha}}\right)=\Phi(\mu_{\alpha},\sigma_{\alpha}),\label{eq:mean_field_maintext}
\end{eqnarray}
where the fluctuations of the input $h_{\alpha}$ to a neuron in population
$\alpha$ are to good approximation Gaussian with the moments
\begin{eqnarray}
\mu_{\alpha}=\langle h_{\alpha}\rangle & = & \sum_{\beta}K_{\alpha\beta}J_{\alpha\beta}m_{\beta}\label{eq:mu_sigma_input}\\
\sigma_{\alpha}^{2}=\langle\delta h_{\alpha}^{2}\rangle & = & \sum_{\beta}K_{\alpha\beta}J_{\alpha\beta}^{2}m_{\beta}(1-m_{\beta}).\nonumber 
\end{eqnarray}
To determine the average activities in the network, the mean-field
equation \eqref{eq:mean_field_maintext} needs to be solved self-consistently,
as the right-hand side depends on the mean activities $m_{\alpha}$
through \eqref{eq:mu_sigma_input}, as explained in \nameref{sub:self_consistent_mf_corr}.
Here $K_{\alpha\beta}$ denotes the number of connections from population
$\beta$ to $\alpha$, and $J_{\alpha\beta}$ their average synaptic
amplitude. Once the mean activity in the network has been found, we
can determine the structure of correlations. For simplicity we focus
on the zero time lag correlation, $c_{ij}=\langle\delta n_{i}(t)\delta n_{j}(t)\rangle_{t}$,
where $\delta n_{i}(t)=n_{i}(t)-\langle n_{i}\rangle_{t}$ is the
deflection of neuron $i$'s activity from baseline and $a_{i}=\langle\delta n_{i}^{2}(t)\rangle_{t}=\langle n_{i}\rangle_{t}(1-\langle n_{i}\rangle_{t})$
is the variance of neuron $i$'s activity. Starting from the master
equation for the network of binary neurons, in \nameref{sec:Methods}
for completeness and consistency in notation we re-derive the self-consistent
equation that connects the cross covariances $c_{\alpha\beta}$ averaged
over pairs of neurons from population $\alpha$ and $\beta$ and the
variances $a_{\alpha}$ averaged over neurons from population $\alpha$
\begin{eqnarray}
c_{\alpha\beta} & = & \frac{1}{N_{\alpha}N_{\beta}}\sum_{k\in\alpha,l\in\beta,k\neq l}\, c_{kl}\label{eq:pop_avg_crosscorr}\\
a_{\alpha} & = & \frac{1}{N_{\alpha}}\sum_{k\in\alpha}a_{k}.\nonumber 
\end{eqnarray}
The obtained inhomogeneous system of linear equations \eqref{eq:corrected_corr_structure}
reads \cite{Ginzburg94}

\begin{eqnarray}
2c_{\alpha\beta} & = & \frac{1}{N_{\beta}}w_{\alpha\beta}a_{\beta}+\sum_{\gamma\in\{E,I,x\}}w_{\alpha\gamma}c_{\gamma\beta}+\text{transpose}(\alpha\leftrightarrow\beta).\label{eq:zero_time_lag_corr}
\end{eqnarray}
Here $w_{\alpha\beta}=S(\mu_{\alpha},\sigma_{\alpha})\, K_{\alpha\beta}\, J_{\alpha\beta}$
measures the effective linearized coupling strength from population
$\beta$ to population $\alpha$. It depends on the number of connections
$K_{\alpha\beta}$ from population $\beta$ to $\alpha$, their average
synaptic amplitude $J_{\alpha\beta}$ and the susceptibility $S_{\alpha}$
of neurons in population $\alpha$. The susceptibility $S(\mu_{\alpha},\sigma_{\alpha})$
given by \eqref{eq:susceptibility} quantifies the influence a fluctuation
in the input to a neuron in population $\alpha$ has on the output.
It depends on the working point $(\mu_{\alpha},\sigma_{\alpha})$
of the neurons in population $\alpha$. The autocorrelations $a_{E}$,
$a_{I}$ and $a_{X}$ are the inhomogeneity in the system of equations,
so they drive the correlations, as pointed out earlier \cite{Ginzburg94}.
This is in line with the linear theories \cite{Tetzlaff12_e1002596,Helias13_023002}
for leaky integrate-and-fire model neurons, where cross-correlations
are proportional to the auto-correlations. The system of equations
\eqref{eq:zero_time_lag_corr} is identical to \cite[ eqs. (9.14)-(9.16)]{Ginzburg94}.
Note that this description holds for finite-sized networks. With the
symmetry $c_{EI}=c_{IE}$, \eqref{eq:zero_time_lag_corr} can be written
in matrix form as
\begin{eqnarray}
A\left(\begin{array}{c}
c_{EE}\\
c_{EI}\\
c_{II}
\end{array}\right) & = & B\,\left(\begin{array}{c}
\frac{a_{E}}{N_{E}}\\
\frac{a_{I}}{N_{I}}
\end{array}\right)+C\,\left(\begin{array}{c}
c_{EX}\\
c_{IX}
\end{array}\right)\label{eq:corrected_corr_structure}\\
D\left(\begin{array}{c}
c_{EX}\\
c_{IX}
\end{array}\right) & = & E\,\frac{a_{X}}{N_{X}}.\nonumber 
\end{eqnarray}
The explicit forms of the matrices $A,\ldots,E$ are given in \eqref{eq:matrix_form_pop_correlation}.
This system of linear equations can be solved by elementary methods.
From the structure of the equations it follows, that the correlations
between the external input and the activity in the network, $c_{EX}$
and $c_{IX}$, are independent of the other correlations in the network.
They are solely determined by the solution of the system of equations
in the second line of \eqref{eq:corrected_corr_structure}, driven
by the fluctuations of the external drive $a_{X}/N_{X}$. The correlations
among the neurons within the network are given by the solution of
the first system in \eqref{eq:corrected_corr_structure}. They are
hence driven by two terms, the fluctuations of the neurons within
the network proportional to $a_{E}/N_{E}$ and $a_{I}/N_{I}$ and
the correlations between the external population and the neurons in
the network, $c_{EX}$ and $c_{IX}$.

The second line of \eqref{eq:corrected_corr_structure} shows that
all correlations depend on the size $N_{X}$ of the external population.
Since the number $K=pN$ of randomly drawn afferents per neuron from
this population is constant, the mean number of shared inputs to a
pair of neurons is $K^{2}/N_{X}$. In the extreme case $N_{X}=K$
on the left of \prettyref{fig:correlations_ext} all neurons receive
exactly identical input. If the recurrent connectivity would be absent,
we would hence have perfectly correlated activity within the local
network, the covariance between two neurons would be equal to their
variance $a_{\alpha}=m_{\alpha}(1-m_{\alpha})$, in this particular
network $a_{\alpha}\simeq0.25$. \prettyref{fig:correlations_ext}A
shows that the covariance in the recurrent network is much smaller;
on the order of $10^{-4}$. The reason is the recently reported mechanism
of decorrelation \cite{Renart10_587}, explained by the negative feedback
in inhibition-dominated networks \cite{Tetzlaff12_e1002596}. Increasing
the size of the external population decreases the amount of shared
input, as seen in \prettyref{fig:correlations_ext}C. In the limit
where the external drive is replaced by a constant value (shown as
point ``$\text{DC}$''), the external drive does consequently not
contribute to correlations in the network. \prettyref{fig:correlations_ext}A
shows that the relative position of the three curves does not change
with $N_{X}$. The overall offset, however, changes. This can be understood
by inspecting the analytical result \eqref{eq:corrected_corr_structure}:
The solution of this system of linear equations is a superposition
of two contributions. One is due to the externally imposed fluctuations,
proportional to $a_{X}/N_{X}$, the other is due to fluctuations generated
within the local network, proportional to $a_{E}/N_{E}$ and $a_{I}/N_{I}$.
Varying the size of the external population only changes the external
contribution, causing the variation in the offset, while the internal
contribution, causing the splitting between the three curves, remains
constant. In the extreme case $a_{X}=0$ ($\text{DC input}$), we
still observe a similar structure. The slightly larger splitting is
due to the reduced variance $\sigma_{\alpha}^{2}$ in the single neuron
input, which consequently increases the susceptibility $S_{\alpha}$
\eqref{eq:susceptibility}. 

\prettyref{fig:correlations_ext}D shows the probability distribution
of the input $h_{\alpha}$ to a neuron in population $\alpha=E$.
The histogram is well approximated by a Gaussian. The first two moments
of this Gaussian are $\mu_{\alpha}$ and $\sigma_{\alpha}^{2}$ given
by \eqref{eq:mu_sigma_input}, if correlations among the afferents
are neglected. This approximation deviates from the result of direct
simulation. Taking the correlations among the afferents into account
affects the variance in the input according to \eqref{eq:variance_corr}.
The latter approximation is a better estimate of the input statistics,
as shown in \prettyref{fig:correlations_ext}D. This improved estimate
can be accounted for in the solution of the mean-field equation \eqref{eq:mean_field_maintext},
which in turn affects the correlations via the susceptibility $S_{\alpha}$.
Iterating this procedure until convergence, as explained in \nameref{sub:self_consistent_mf_corr},
yields the semi-analytical results presented in \prettyref{fig:correlations_ext}.

\subsection{Cancellation of input correlations\label{sub:cancellation_input}}

\begin{figure}
\noindent \begin{centering}
\includegraphics[scale=0.9]{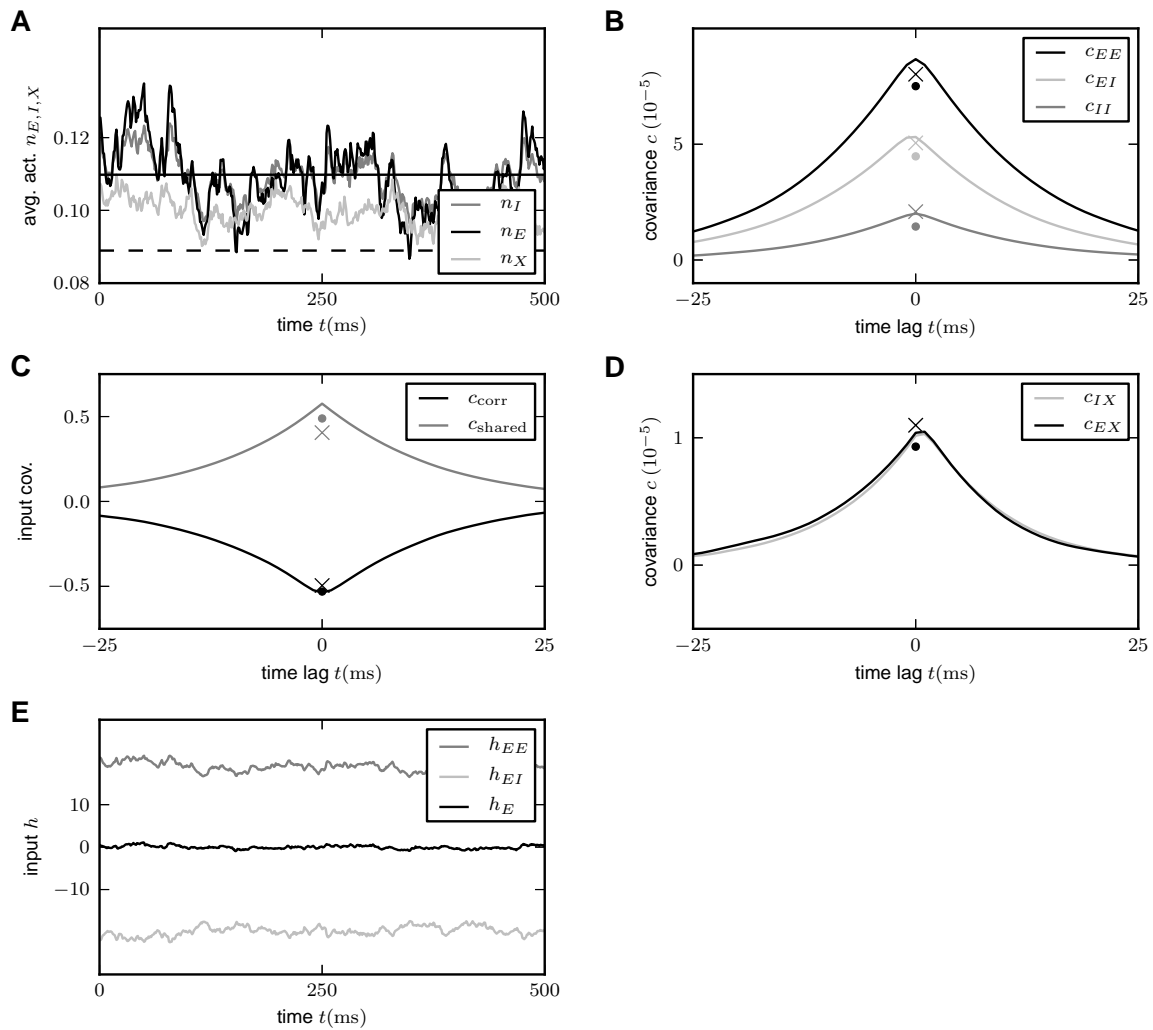}
\par\end{centering}

\caption{Activity in a network of $3N=3\times8192$ binary neurons with synaptic
amplitudes $J_{\alpha E}=J_{\alpha X}=5/\sqrt{N}$, $J_{\alpha I}=-10/\sqrt{N}$
depending exclusively on the type of the sending neuron ($E$ or $I$).
Each neuron receives $K=pN$ randomly drawn inputs (fixed in-degree,
$p=0.2$). \textbf{A} Population averaged activity (black $E$, gray
$I$, light gray $X$). Analytical prediction \prettyref{eq:cancellation_mu}
for the mean activities $m_{E}=m_{I}$ (dashed horizontal line) and
numerical solution of mean field equation \eqref{eq:mean_field_activity}
(solid horizontal line). \textbf{B} Cross covariance between excitatory
neurons (black), between inhibitory neurons (gray), and between excitatory
and inhibitory neurons (light gray). Theoretical results \prettyref{eq:corrected_corr_structure}
shown as dots. St. Andrew's Crosses indicate the theoretical prediction
of leading order in $N^{-1}$ \eqref{eq:cov_symm_infinite}. \textbf{C}
Correlation between the input currents to a pair of excitatory neurons.
The black curve is the contribution due to pairwise correlations $c_{\mathrm{corr}}$,
the gray curve is the contribution of shared input $c_{\mathrm{shared}}$.
The symbols show the theoretical expectation \prettyref{eq:c_shared}
and \prettyref{eq:c_corr} based on \eqref{eq:cov_symm_infinite}
(crosses) and based on \prettyref{eq:corrected_corr_structure} (dots).
\textbf{D} Similar to B, but showing the correlations between external
neurons and neurons in the excitatory and inhibitory population. Note
that both theories yield $c_{EX}=c_{IX}$, so for each theory (\eqref{eq:cov_symm_infinite}
crosses, \prettyref{eq:corrected_corr_structure} dots) only the symbol
for $c_{EX}$ is visible. \textbf{E} Contributions $h_{EE}$ (gray)
due to excitatory synapses and $h_{EI}$ (light gray) due to inhibitory
synapses to the input $h_{E}$ averaged over all excitatory neurons.
Duration of simulation $T=100\mathrm{\, s}$, mean activity $m_{X}=0.1$,
$m_{E}\simeq m_{I}\simeq0.11$, other parameters as in \prettyref{fig:correlations_ext}.\foreignlanguage{english}{\label{fig:symmetric}}}
\end{figure}

We would like to understand how the structure of correlations relates
to the earlier report of fast tracking \cite{Vreeswijk96,VanVreeswijk98_1321}.
Small correlations observed in balanced recurrent networks were explained
by the property of recurrent networks to track their input on a fast
time-scale \cite[their eq. (2)]{Renart10_587}. \prettyref{fig:symmetric}A
shows the population activities in a network of three populations
for fixed numbers of neurons $N_{x}=N_{E}=N_{I}=N$ and symmetric
connectivity as in \prettyref{fig:correlations_ext}. The deflections
of the excitatory and the inhibitory population partly resemble those
of the external drive to the network, but partly the fluctuations
seem to be independent. Our theoretical result for the correlation
structure explains this result \eqref{eq:corrected_corr_structure}:
the fluctuations in the network are not only driven by external input
(proportional to $a_{X}$), but are also driven by the fluctuations
generated within the local populations (proportional to $a_{E}$ and
$a_{I}$). The idea of fast tracking of the external signal was derived
from the observation that the fluctuations in the population-averaged
input $h_{\alpha}=\frac{1}{N_{\alpha}}\sum_{i\in\alpha}h_{i}$ are
suppressed \cite{Renart10_587}. This suppression can be observed
by decomposing the input $h_{\alpha}$ to the population $\alpha$
into contributions from excitatory (including external neurons) and
from inhibitory cells, $h_{\alpha E}=(KJ)_{\alpha E}n_{E}+(KJ)_{\alpha X}n_{X}$
and $h_{\alpha I}=(KJ)_{\alpha I}n_{I}$, respectively, where we used
the short hand $(KJ)_{\alpha\beta}=K_{\alpha\beta}J_{\alpha\beta}$.
As shown in \prettyref{fig:symmetric}E, the contributions of excitation
and inhibition cancel each other so that the total input fluctuates
close to the threshold (here $\theta=1$) of the neurons: the network
is in the balanced state \cite{Vreeswijk96}. Moreover, this cancellation
not only holds for the mean value, but also for fast fluctuations,
which are consequently reduced in the sum $h_{\alpha}$ compared to
the individual components $h_{\alpha E}$ and $h_{\alpha I}$ (\prettyref{fig:symmetric}E).
We will now show that this suppression of fluctuations directly implies
a relation for the correlation $\langle\delta h_{i}\delta h_{j}\rangle$
between the inputs to a pair $(i,j)$ of individual neurons. There
are two distinct contributions to this correlation $\langle\delta h_{i}\delta h_{j}\rangle=c_{\mathrm{shared},\alpha}+c_{\mathrm{corr},\alpha}$,
one due to common inputs shared by the pair of neurons (both neurons
$i,j$ assumed to belong to population $\alpha$) 
\begin{eqnarray}
c_{\mathrm{shared},\alpha} & = & \sum_{\beta\in\{E,I,X\}}(KJ)_{\alpha\beta}^{2}\frac{a_{\beta}}{N_{\beta}}\label{eq:c_shared}
\end{eqnarray}
 and one due to the correlations between afferents 
\begin{eqnarray}
c_{\mathrm{corr},\alpha} & = & \sum_{\beta,\gamma\in\{E,I,X\}}(KJ)_{\alpha\beta}(KJ)_{\alpha\gamma}c_{\beta\gamma}.\label{eq:c_corr}
\end{eqnarray}
\prettyref{fig:symmetric}C shows these two contributions to be of
opposite sign but approximately same magnitude. \prettyref{fig:correlations_ext}C
shows a further decomposition of the input correlation into contributions
due to the external sources and due to connections from within the
local network. The sum of all components is much smaller than each
individual component. This cancellation is equivalent to  small fluctuations
in the population-averaged input $\langle\delta h_{\alpha}^{2}\rangle\simeq0$,
because 
\begin{eqnarray}
0\simeq\langle\delta h_{\alpha}^{2}\rangle=\left\langle \left(\sum_{\beta\in\{E,I,X\}}(KJ)_{\alpha\beta}\delta n_{\beta}\right)^{2}\right\rangle  & = & \sum_{\beta,\gamma\in\{E,I,X\}}(KJ)_{\alpha\beta}(KJ)_{\alpha\gamma}\langle\delta n_{\beta}\delta n_{\gamma}\rangle\label{eq:cancellation_input}\\
 & = & \sum_{\beta\in\{E,I,X\}}(KJ)_{\alpha\beta}^{2}\,\frac{a_{\beta}}{N_{\beta}}+\sum_{\beta,\gamma\in\{E,I,X\}}(KJ)_{\alpha\beta}(KJ)_{\alpha\gamma}\, c_{\beta\gamma}\nonumber \\
 & = & c_{\mathrm{shared,\alpha}}+c_{\mathrm{corr},\alpha},\nonumber 
\end{eqnarray}
where in the second step we used the general relation between the
covariance $\langle\delta n_{\beta}\delta n_{\gamma}\rangle$ among
two population averaged signals $n_{\beta}$ and $n_{\gamma}$, the
population-averaged variance $a_{\beta}$, and the pairwise averaged
covariances $c_{\beta\gamma}$, which reads \cite[cf. eq. (1)]{Tetzlaff12_e1002596}
\begin{eqnarray}
\langle\delta n_{\beta}\delta n_{\gamma}\rangle & = & \left\langle \frac{1}{N_{\beta}N_{\gamma}}\sum_{i\in\beta,j\in\gamma}\delta n_{i}\delta n_{j}\right\rangle \nonumber \\
 & = & \delta_{\beta\gamma}\frac{1}{N_{\beta}^{2}}\sum_{i\in\beta}\langle\delta n_{i}^{2}\rangle+\frac{1}{N_{\beta}N_{\gamma}}\sum_{i\in\beta,j\in\gamma,i\neq j}\langle\delta n_{i}\delta n_{j}\rangle\label{eq:pop_fluctuation-1}\\
 & = & \delta_{\beta\gamma}\frac{1}{N_{\beta}}a_{\beta}+c_{\beta\gamma}.\nonumber 
\end{eqnarray}

\begin{figure}
\begin{centering}
\includegraphics[scale=0.9]{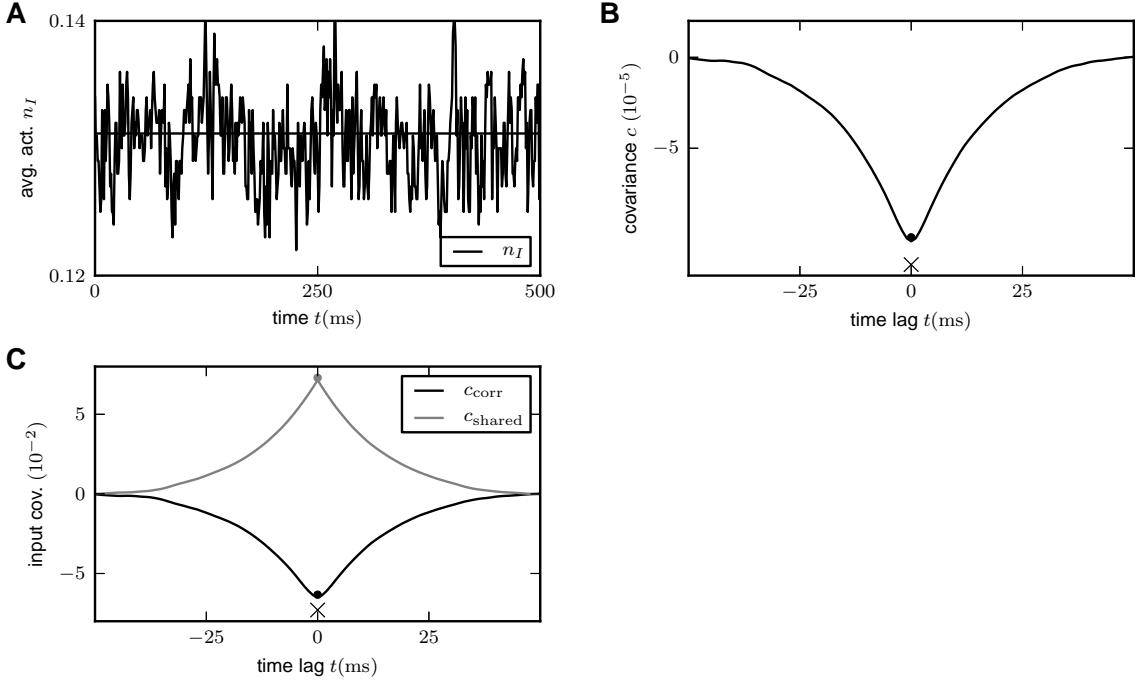}
\par\end{centering}

\caption{Suppression of correlations by purely inhibitory feedback in absence
of external fluctuations. Activity in a network of $N=1000$ binary
inhibitory neurons with synaptic amplitudes $J=-\frac{8}{\sqrt{N}}$.
Each neuron receives $K=pN$ randomly drawn inputs (fixed in-degree)
with $p=0.1$. \textbf{A} Population averaged activity. Numerical
solution of mean field equation \eqref{eq:mean_field_activity} (solid
horizontal line). \textbf{B} Cross covariance between inhibitory neurons.
Theoretical result \eqref{eq:corr_inh} shown as dot. St. Andrew's
Cross indicates the leading order term $c=-\frac{a}{N}$. \textbf{C}
Correlation between the input currents to a pair of excitatory neurons.
The black curve is the contribution due to pairwise correlations $c_{\mathrm{corr}}$,
the gray curve is the contribution of shared input $c_{\mathrm{shared}}$.
The dot symbols show the theoretical expectations \prettyref{eq:corr_input_inh}
based on the leading order (crosses) and based on the full solution
\prettyref{eq:corr_inh} (dot). Threshold of neurons $\theta=\frac{1}{10}pNJ+\frac{J}{2}$.\label{fig:inh-suppression}}
\end{figure}
This suppression of fluctuations in the population-averaged input
is a consequence of the overall negative feedback in these networks:
a fluctuation $\delta h_{\alpha}$ of the population averaged input
$h_{\alpha}$ causes a response in network activity which is coupled
back with a negative sign, counteracting its own cause and hence suppressing
the fluctuation $\delta h_{\alpha}$. Expression \eqref{eq:cancellation_input}
is an algebraic identity showing that hence also correlations between
the total inputs to a pair of cells must be suppressed. Qualitatively
this property can be understood by inspecting the mean-field equation
\eqref{eq:mean_field_activity} for the population-averaged activities,
where we linearized the gain function $\Phi$ around the stationary
mean-field solution to obtain

\begin{eqnarray}
\tau\frac{d}{dt}\left(\begin{array}{c}
\delta n_{E}\\
\delta n_{I}
\end{array}\right)-\left(\begin{array}{c}
\delta n_{E}\\
\delta n_{I}
\end{array}\right) & = & \left(\begin{array}{cc}
w_{EE} & w_{EI}\\
w_{IE} & w_{II}
\end{array}\right)\left(\begin{array}{c}
\delta n_{E}\\
\delta n_{I}
\end{array}\right)+\text{noise}\label{eq:mean_field_fluct-1}\\
\text{with }w_{\alpha\beta} & = & S(\mu_{\alpha},\sigma_{\alpha})\,(KJ)_{\alpha\beta}\nonumber \\
\text{and }S(\mu_{\alpha},\sigma_{\alpha}) & = & \frac{\partial\Phi(\mu_{\alpha},\sigma_{\alpha})}{\partial\mu_{\alpha}}.\nonumber 
\end{eqnarray}
Here the noise term qualitatively describes the fluctuations caused
by the stochastic update process and the external drive. After transformation
into the coordinate system of eigenvectors $\u_{i}$ (with eigenvalue
$\lambda_{i}$) of the effective connectivity matrix $\W$, each component
fulfills the differential equation
\begin{eqnarray*}
\tau\frac{d}{dt}\delta\u_{i}(t)+\delta\u_{i}(t) & =\lambda_{i}\delta\u_{i}(t)+ & \text{projection of noise on direction \ensuremath{\u_{i}}}.
\end{eqnarray*}
For stability the eigenvalues must be $\Re(\lambda_{i})<1$. In the
example of the homogeneous $E-I$ network we have one negative eigenvalue
$\lambda_{2}=SKJ(1-\gamma g)<0$. The fluctuations $\delta\u_{2}$
are hence suppressed in this direction so the contribution $\delta\mathbf{h}_{2}=\W\delta\u_{2}$
to the fluctuations on the input side is small. The other eigenvalue
is $\lambda_{1}=0$, so fluctuations are only mildly suppressed in
direction $\delta\u_{1}$. However, on the input side of the neurons,
these fluctuations are not seen, since their contribution to the input
field is by the vanishing eigenvalue $\delta\mathbf{h}_{1}=\W\delta\u_{1}=0$.
This explains why fluctuations of $\delta h_{\alpha}$ are always
small in networks stabilized by inhibition-dominated negative feedback.
This argument also shows why the suppression of input-correlations
does not rely on a balance between excitation and inhibition; it is
as well observed in purely inhibitory networks \cite[cf. text following eq. (21) therein]{Tetzlaff12_e1002596},
where the overall negative feedback suppresses population fluctuations
$\delta h_{\alpha}$ in exactly the same manner, as the only appearing
eigenvalue in this case is negative. \prettyref{fig:inh-suppression}
shows the correlations in a purely inhibitory network without any
external fluctuations. In this network the neurons are autonomously
active due to a negative threshold $\theta$, which, by the cancellation
argument $\langle h\rangle\simeq\theta$, was chosen to obtain a mean
activity of about $0.1$. Pairwise correlations follow from \eqref{eq:zero_time_lag_corr}
to be negative, 
\begin{align}
c & =\frac{w}{1-w}\frac{a}{N}<0\label{eq:corr_inh}
\end{align}
 and approach $c=-\frac{a}{N}$ in the limit of strong coupling. Hence
the contributions to the input correlation follow from \eqref{eq:c_shared}
and \eqref{eq:c_corr} as 
\begin{align}
c_{\mathrm{corr}} & =(KJ)^{2}c=(KJ)^{2}\frac{w}{1-w}\,\frac{a}{N}\label{eq:corr_input_inh}\\
c_{\mathrm{shared}} & =(KJ){}^{2}\,\frac{a}{N},\nonumber 
\end{align}
so that for strong negative feedback $|w|\gg1$ the contribution due
to correlations approaches $c_{\mathrm{corr}}\to-(KJ)^{2}\frac{a}{N}=-c_{\mathrm{shared}}$.
In this limit the two contributions cancel each other as in the inhibition-dominated
network with excitation and inhibition. For finite coupling $|c_{\mathrm{shared}}|>|c_{\mathrm{corr}}|$,
so the total currents are always positively correlated.

\begin{figure}
\noindent \begin{centering}
\includegraphics[scale=0.9]{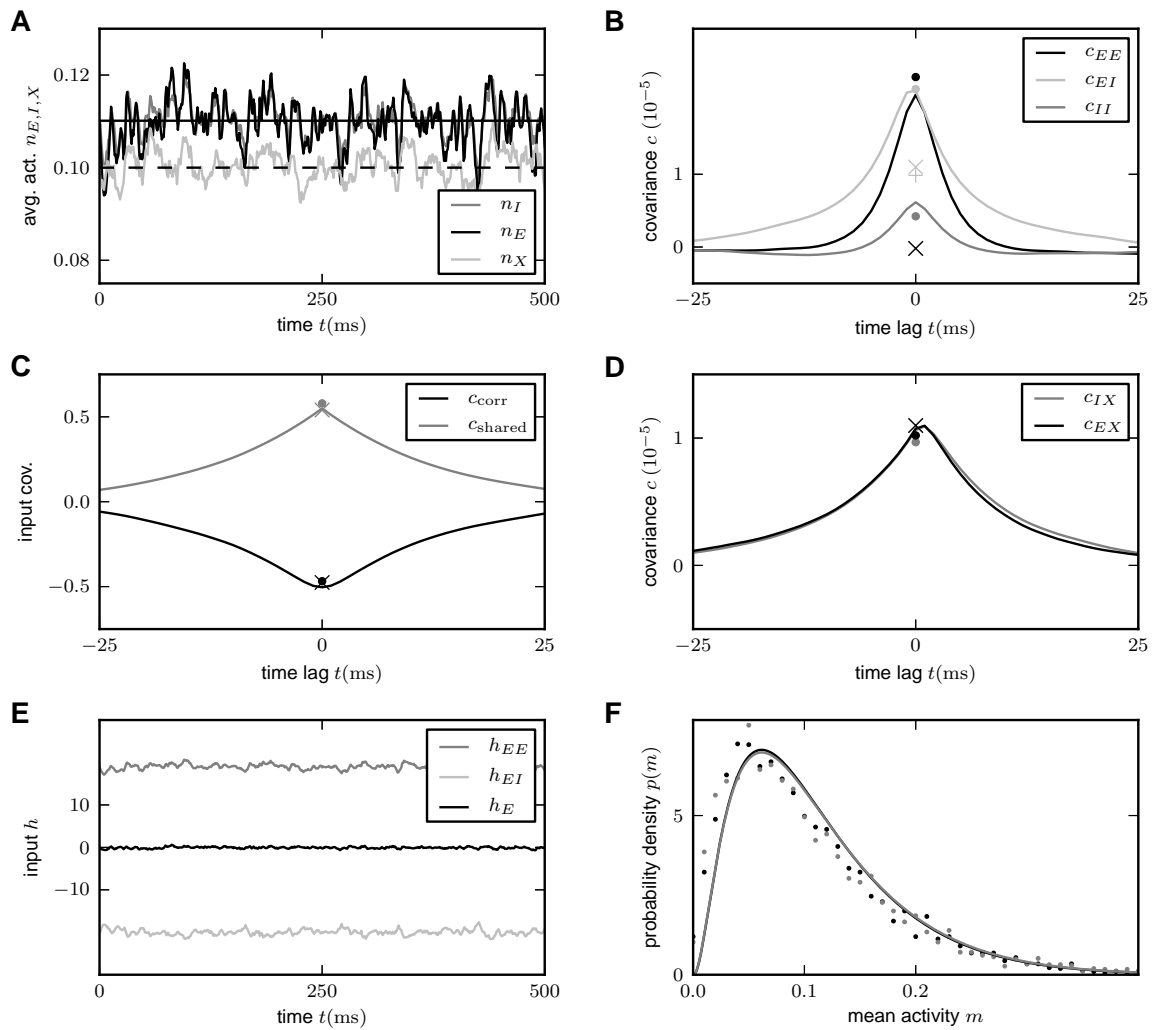}
\par\end{centering}

\caption{Activity in a network of $3N=3\times8192$ binary neurons as described
in \cite{Renart10_587}, with $J_{EE}=5/\sqrt{N}$, $J_{EI}=-10/\sqrt{N}$,
$J_{IE}=5/\sqrt{N}$, $J_{II}=-9/\sqrt{N}$, $J_{EX}=5/\sqrt{N}$,
$J_{IX}=4/\sqrt{N}$. Number $K$ of synaptic inputs binomially distributed
as $K\sim B(N,p)$, with connection probability $p=0.2$. \textbf{A}
Population averaged activity (black $E$, gray $I$, light gray $X$).
Analytical prediction \prettyref{eq:cancellation_mu} for the mean
activities $m_{E}=m_{I}$ (dashed horizontal line) and numerical solution
of mean field equation \eqref{eq:mean_field_activity} (solid horizontal
line). \textbf{B} Cross correlation between excitatory neurons (black
curve), between inhibitory neurons (gray curve), and between excitatory
and inhibitory neurons (light gray curve) obtained from simulation.
St. Andrew's Crosses show the theoretical prediction from \cite[suppl. eqs. 38,39]{Renart10_587}
(prediction yields $c_{EE}\simeq c_{II}\simeq-2\,10^{-7}$, so only
one cross is visible). Dots show the theoretical prediction \prettyref{eq:corrected_corr_structure}.
The plus symbol shows the prediction for the correlation $c_{EI}$
when terms proportional to $a_{E}$ and $a_{I}$ are set to zero.
\textbf{C} Correlation between the input currents to a pair of excitatory
neurons. Contribution due to pairwise correlations $c{}_{\mathrm{corr},E}$
(black curve) and due to shared input $c_{\mathrm{shared},E}$ (gray
curve). Symbols show the theoretical predictions based on \cite{Renart10_587}
(crosses) and based on \prettyref{eq:corrected_corr_structure} (dots).
\textbf{D} Similar to B, but showing the correlations between external
neurons and neurons in the excitatory and inhibitory population. \textbf{E}
Fluctuating input $h_{E}$ averaged over the excitatory population
(black), separated into contributions from excitatory synapses $h_{EE}$
(gray) and from inhibitory synapses $h_{EI}$ (light gray). \textbf{F}
Distribution of time averaged activity obtained by direct simulation
(symbols) and analytical prediction \eqref{eq:rate_distribution}
using the numerically evaluated self-consistent solution for the first
$m_{E}\simeq m_{I}\simeq0.11$ and second moments $q_{E}\simeq0.019$,
$q_{I}\simeq0.018$ \eqref{eq:second_moment_rate_distribution}.
Duration of simulation $T=100\mathrm{\, s}$, mean activity $m_{X}=0.1$,
other parameters as in \prettyref{fig:correlations_ext}.\label{fig:renart_binary}}
\end{figure}

It is easy to see that the cancellation condition \prettyref{eq:cancellation_input}
does not uniquely determine the structure of correlations in an $E-I$
network. \prettyref{fig:renart_binary} shows the same measures of
activity as \prettyref{fig:symmetric}, but for the network connectivity
used in \cite[ their Fig. 2]{Renart10_587}. Except for $J_{II}$
and $J_{IX}$ the parameters are the same as in \prettyref{fig:symmetric}.
Moreover, we distributed the number of incoming connections $K$ per
neuron according to a binomial distribution as in the original publication.
As before, the cancellation of fluctuations on the input side is
evident from \prettyref{fig:renart_binary}E, which is here realized
by a different structure of covariances $c_{EE}\simeq c_{EI}>c_{II}$,
shown in \prettyref{fig:renart_binary}B.  The structure of correlations
in a finite network is not uniquely determined by $\langle\delta h\rangle=0$,
as seen in \prettyref{fig:renart_binary}B. As an example consider
the correlation structure predicted in the limit of infinite network
size \cite[Supplementary material, eqs. 38-39]{Renart10_587}, which
also fulfills $\langle\delta h\rangle=0$, but does not coincide with
the results obtained by direct simulation of the finite network. By
construction and by virtue of \eqref{eq:cancellation_input} this
correlation structure, however, still fulfills the cancellation condition
on the input side, as visualized in \prettyref{fig:renart_binary}C.
We show in \nameref{sub:Limit-of-infinite} below that this is due
to the theory being valid only in the limit of infinite network size,
neglecting the contribution of fluctuations of the local populations
($E$,$I$), as they appear in \eqref{eq:corrected_corr_structure}.
Formally this is apparent from \cite[eq. (2)]{Renart10_587}, stating
that fluctuations are predominantly caused by the external input,
reflected in the expression $c_{EI}\propto a_{X}$. This can be demonstrated
explicitly by setting $a_{E}=0$ and $a_{I}=0$ in \prettyref{eq:corrected_corr_structure},
resulting in a similar prediction for $c_{EI}$, as shown in \prettyref{fig:renart_binary}B
(plus symbol). The remaining deviation between the theories is due
to the different susceptibilities $S$ used by the two approaches.
Note that the full theory \prettyref{eq:corrected_corr_structure}
predicts the structure of correlations with high accuracy. In summary,
the cancellation condition imposes a constraint on the structure of
correlations but is not sufficient as a unique determinant.

The distribution of the in-degree is an additional source of variability.
It causes a distribution of the mean activity of the neurons in the
network, as shown in \prettyref{fig:renart_binary}F. The shape of
the distribution can be assessed analytically by self-consistently
solving a system of equations for the first $m_{\alpha}$ \eqref{eq:first_moment_rate_distribution}
and second moment $q_{\alpha}$ \eqref{eq:second_moment_rate_distribution}
of the rate distribution \cite{VanVreeswijk98_1321}, as described
in \nameref{sec:Influence-of-inhomogeneity}. The resulting second
moments $q_{E}\simeq0.0185$ ($0.0175$ by simulation) and $q_{I}\simeq0.0184$
($0.0180$ by simulation) are small compared to the mean activity
$m_{E}\simeq m_{I}\simeq0.11\ll1$. For the prediction of the covariances
shown in \prettyref{fig:renart_binary}B-D we employed the semi-analytical
self-consistent solution to determine the variances $a_{\alpha}=m_{\alpha}-q_{\alpha}$.
The difference to the approximate value $a_{\alpha}\simeq m_{\alpha}(1-m_{\alpha})<m_{\alpha}-[m_{\alpha}^{2}]$
is, however, small for low mean activity.

\subsection{Limit of infinite network size\label{sub:Limit-of-infinite}}

To relate the finite-size correlations presented in the previous sections
to earlier studies of the dominant contribution to the correlations
in the limit of infinitely large networks \cite{Renart10_587}, we
here take the limit $N\to\infty$. For non-homogeneous connectivity,
we recover the earlier result \cite{Renart10_587} in \nameref{sub:Inhomogeneous-connectivity}.
In \nameref{sub:Symmetric-connectivity} we show that the correlations
converge to a different limit than what would be expected from the
idea of fast tracking.

Starting from \eqref{eq:population_correlation} we follow \cite[Supplementary]{Renart10_587}
and introduce the covariances between population-averaged activities
as $r_{\alpha\beta}=c_{\alpha\beta}+\delta_{\alpha\beta}\frac{a_{\alpha}}{N_{\alpha}}$,
which leads to

\begin{align}
2(r_{\alpha\beta}-\delta_{\alpha\beta}\frac{a_{\alpha}}{N_{\alpha}}) & =\sum_{\gamma\in\{E,I,X\}}\left(w_{\alpha\gamma}r_{\gamma\beta}+w_{\beta\gamma}r_{\gamma\alpha}\right)\nonumber \\
\sum_{\gamma\in\{E,I,X\}}\left(\underbrace{(\delta_{\alpha\gamma}-w_{\alpha\gamma})}_{\equiv m_{\alpha\gamma}}r_{\gamma\beta}+\underbrace{(\delta_{\beta\gamma}-w_{\beta\gamma})}_{\equiv m_{\beta\gamma}}r_{\gamma\alpha}\right) & =2\delta_{\alpha\beta}\frac{a_{\alpha}}{N_{\alpha}}\label{eq:cov_renart_r}\\
MR+(MR)^{T} & =2\mathrm{diag}(\{\frac{a_{\alpha}}{N_{\alpha}}\}).\nonumber 
\end{align}
The general solution of the continuous Lyapunov equation stated in
the last line can be obtained by projecting onto the set of left-sided
eigenvectors of $M$ (see e.g. \cite{Ginzburg94} eq.~6.14). 
Alternatively the system of linear equations \eqref{eq:cov_renart_r}
may be written explicitly as

\begin{align}
\underbrace{\left(\begin{array}{ccc}
2-2w_{EE} & -2w_{EI} & 0\\
-w_{IE} & 2-\left(w_{EE}+w_{II}\right) & -w_{EI}\\
0 & -2w_{IE} & 2-2w_{II}
\end{array}\right)}_{\equiv\tilde{M}}\left(\begin{array}{c}
r_{EE}\\
r_{EI}\\
r_{II}
\end{array}\right) & =\left(\begin{array}{cc}
2w_{EX} & 0\\
w_{IX} & w_{EX}\\
0 & 2w_{IX}
\end{array}\right)\left(\begin{array}{c}
r_{EX}\\
r_{IX}
\end{array}\right)+2\left(\begin{array}{c}
\frac{a_{E}}{N_{E}}\\
0\\
\frac{a_{I}}{N_{I}}
\end{array}\right)\label{eq:cov_split_ext_intri}\\
\left(\begin{array}{cc}
2-w_{EE} & -w_{EI}\\
w_{IE} & 2-w_{II}
\end{array}\right)\left(\begin{array}{c}
r_{EX}\\
r_{IX}
\end{array}\right) & =\left(\begin{array}{c}
w_{EX}\\
w_{IX}
\end{array}\right)\frac{a_{X}}{N_{X}}.\nonumber 
\end{align}
The solution of the latter equation is given by \eqref{eq:external_correlation},
so $r_{\alpha X}\propto\frac{a_{X}}{N_{X}}$. We observe that the
right hand side of the first line in \eqref{eq:cov_split_ext_intri}
contains again two source terms, those corresponding to fluctuations
caused by the external drive (proportional to $r_{\alpha X}\propto\frac{a_{X}}{N_{X}}$)
and those due to fluctuations generated within the network (proportional
to $a_{E}$ or $a_{I}$). This motivates our definition of the two
contributions $r_{\alpha\beta}^{\mathrm{ext.}}$ and $r_{\alpha\beta}^{\mathrm{int.}}$
as
\begin{align}
\tilde{M\,}\left(\begin{array}{c}
r_{EE}^{\mathrm{ext.}}\\
r_{EI}^{\mathrm{ext.}}\\
r_{II}^{\mathrm{ext.}}
\end{array}\right) & =\left(\begin{array}{cc}
2w_{EX} & 0\\
w_{IX} & w_{EX}\\
0 & 2w_{IX}
\end{array}\right)\left(\begin{array}{c}
r_{EX}\\
r_{IX}
\end{array}\right)\label{eq:r_ext}\\
\tilde{M}\,\left(\begin{array}{c}
r_{EE}^{\mathrm{int.}}\\
r_{EI}^{\mathrm{int.}}\\
r_{II}^{\mathrm{int.}}
\end{array}\right) & =2\left(\begin{array}{c}
\frac{a_{E}}{N_{E}}\\
0\\
\frac{a_{I}}{N_{I}}
\end{array}\right),\label{eq:r_int}
\end{align}
which allows us to write the full solution of \eqref{eq:cov_split_ext_intri}
as $r_{\alpha\beta}=r_{\alpha\beta}^{\mathrm{ext.}}+r_{\alpha\beta}^{\mathrm{int.}}$.
We use the superscripts $\mathrm{ext.}$ and $\mathrm{int.}$ to indicate
the driving sources of the fluctuations coming from outside the network
($\mathrm{ext.}$ driven by $a_{X}$) and coming from within the network
($\mathrm{int.}$ driven by $a_{E}$ and $a_{I}$).

\subsubsection{Inhomogeneous connectivity\label{sub:Inhomogeneous-connectivity}}

\begin{figure}
\begin{centering}
\includegraphics[scale=0.9]{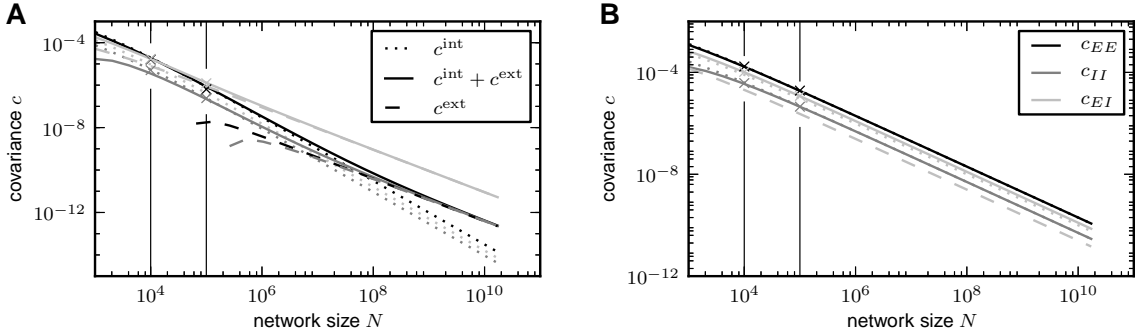}
\par\end{centering}

\caption{Scaling the network size to infinity. Comparison of the solution of
\eqref{eq:corrected_corr_structure} (solid) to the contribution of
the leading order in $1/N$ (dashed). Gray coded are the different
pairs of covariances, black ($c_{EE}$), mid gray ($c_{II}$), light
gray ($c_{EI}$). \textbf{A} Network as in \cite{Renart10_587} with
non-homogeneous synaptic coupling as in \prettyref{fig:renart_binary}.
The dashed curve is given by the leading order term $c_{\alpha\beta}^{\mathrm{ext.}}=r_{\alpha\beta}^{\mathrm{ext.}}-\delta_{\alpha\beta}\frac{a_{\alpha}}{N_{\alpha}}\propto N^{-1}$
\prettyref{eq:cov_renart} and \cite[eqs. (38)-(39)]{Renart10_587}
driven by external fluctuations, the dotted curve is the next order
term $r^{\mathrm{intr.}}\propto N^{-\frac{3}{2}}$ \eqref{eq:r_int},
driven by intrinsic fluctuations generated by the excitatory and inhibitory
population. The dashed curve is not shown for networks smaller than
$\sim10^{6}$ neurons as it assumes negative values. Relative error
of the theory with respect to simulation at $100,000$ neurons is
$73$ percent. The solid curve is the full solution of \eqref{eq:corrected_corr_structure}
$c_{\alpha\beta}=r_{\alpha\beta}^{\mathrm{ext.}}+r_{\alpha\beta}^{\mathrm{intr.}}-\delta_{\alpha\beta}\frac{a_{\alpha}}{N_{\alpha}}$.
The relative error at $100,000$ neurons is $16$ percent. Symbols
show direct simulations. \textbf{B} Network with homogeneous connectivity,
as in \prettyref{fig:symmetric}. Same symbol code as in A. Both contributions
$c_{\alpha\beta}^{\mathrm{ext.}}\propto N^{-1}$ \eqref{eq:r_ext}
and $r^{\mathrm{intr.}}\propto N^{-1}$ \eqref{eq:r_int} show the
same scaling \eqref{eq:c_ext_approx}. Note that for the parameters
here $c_{\alpha\alpha}^{\mathrm{ext.}}\simeq0$, so the only dashed
curve shown is $c_{EI}^{\mathrm{ext.}}$. Symbols indicate the results
of direct simulations; vertical lines are included to guide the eye.\label{fig:Scaling-to-infinity.}}
\end{figure}

In the following we assume inhomogeneous connectivity, meaning that
the synaptic amplitudes not only depend on the type of the sending
neuron but also on the receiving neuron, such that the matrix $\{J_{\alpha\beta}\}$
is invertible. In the limit of large networks with $|w_{\alpha\beta}|\gg1$
the solution \eqref{eq:external_correlation} can be approximated
as

\begin{eqnarray*}
\left(\begin{array}{c}
c_{EX}\\
c_{IX}
\end{array}\right)=\left(\begin{array}{c}
r_{EX}\\
r_{IX}
\end{array}\right) & \simeq & \left(\begin{array}{cc}
w_{EE} & w_{EI}\\
w_{IE} & w_{II}
\end{array}\right)^{-1}\left(\begin{array}{c}
w_{EX}\\
w_{IX}
\end{array}\right)\frac{a_{X}}{N_{X}}\equiv\left(\begin{array}{c}
A_{E}\\
A_{I}
\end{array}\right)\frac{a_{x}}{N_{x}},
\end{eqnarray*}
where the definitions of $A_{E}$ and $A_{I}$ correspond to the ones
of \cite{Renart10_587} if the susceptibility $S$ is the same for
all populations. Solving the first system of equations \eqref{eq:r_ext}
leads to 
\begin{align*}
r_{\alpha\beta}^{\mathrm{ext.}} & \simeq A_{\alpha}A_{\beta}\frac{a_{X}}{N_{X}},
\end{align*}
where we again assumed that $|w_{\alpha\beta}|\gg1$ and therefore
neglected the term $2$ in the sums on the diagonal of the matrix
$\tilde{M}$ \prettyref{eq:cov_split_ext_intri}. Hence the covariance
due to $r_{\alpha\beta}^{\mathrm{ext.}}$ is 
\begin{align}
c_{\alpha\beta}^{\mathrm{ext.}} & =r_{\gamma\beta}^{\mathrm{ext.}}-\delta_{\alpha\beta}\frac{a_{\alpha}}{N_{\alpha}}\label{eq:cov_renart}\\
 & \simeq A_{\alpha}A_{\beta}\frac{a_{X}}{N_{X}}-\delta_{\alpha\beta}\frac{a_{\alpha}}{N_{\alpha}}\propto N^{-1}.\nonumber 
\end{align}
The latter equation is the solution given in \cite[Supplementary, eqs. (38)-(39) ]{Renart10_587}.
The form of the equation shows that this contribution is due to fluctuations
of the population activity driven by the external input, exhibited
by the factor $a_{X}$ driving $r_{\alpha\beta}^{\mathrm{ext.}}$,
where the intrinsic contribution of the single cell autocorrelations
is subtracted. The quantities $A_{E}$ and $A_{I}$ contain the effect
of the recurrence on these externally applied fluctuations and are
independent of network size, so $c^{\mathrm{ext.}}$ decays with $N^{-1}$
as shown in \prettyref{fig:Scaling-to-infinity.}A (dashed curve).

The second contribution $r^{\mathrm{int.}}$ given by the solution
of \eqref{eq:r_int} is driven by the intrinsically generated fluctuations.
As the network tends to infinity, this contribution vanishes faster
than $r^{\mathrm{ext.}}$, because the coupling matrix grows as $\tilde{M}\propto w\propto\sqrt{N}$.
So the term $r^{\mathrm{int.}}$ is a correction to \eqref{eq:cov_renart}
of the order $N^{-\frac{3}{2}}$. This faster decay can be observed
at large network sizes in \prettyref{fig:Scaling-to-infinity.}A (dotted
curve). For finite networks of natural size, however, this term determines
the structure of the correlations. For the parameters chosen in \cite{Renart10_587}
in particular, the contribution $r^{\mathrm{int.}}$ dominates in
networks up to about $10^{7}$ neurons (\prettyref{fig:Scaling-to-infinity.}A).

\subsubsection{Homogeneous connectivity\label{sub:Symmetric-connectivity}}

In the previous section we showed that in agreement with \cite{Renart10_587}
the leading order term $\propto N^{-1}$ dominates the limit of infinitely
large networks and yields practically useful results for random networks
of about $N\ge10^{8}$ neurons. Here we will extend the theory to
homogeneous connectivity, where the synaptic weights only depend on
the type of the sending neuron, i.e. all $J_{\alpha E}=J_{\alpha X}=J$
and $J_{\alpha I}=-gJ$ are the same for all $\alpha$. The matrix
\begin{align}
J\left(\begin{array}{cc}
1 & -g\\
1 & -g
\end{array}\right)\label{eq:eff_conn_symm}
\end{align}
is hence not invertible and the theory in \nameref{sub:Inhomogeneous-connectivity}
not directly applicable. Note that assuming fast tracking here, which
for inhomogeneous connectivity is a consequence of the correlation
structure in the $N\to\infty$ limit \cite[ eq. (2)]{Renart10_587},
due to the degenerate rows of the connectivity here yields
\begin{align}
m_{E}(t)=m_{I}(t) & =A\, m_{X}(t)\nonumber \\
A & =\frac{1}{g-1}.\label{eq:fast_tracking_symm}
\end{align}
Here such an assumption will lead to a wrong result, if $A$ is naively
inserted into equation \eqref{eq:cov_renart} or equivalently into
\cite[Supplementary, eqs. (38)-(39)]{Renart10_587}. In particular,
for the given parameters $g-1=1$ and with the homogeneous activity
(and $a_{x}=a_{E}=a_{I}$) the cross covariances $c_{\alpha\alpha}$
are predicted to approximately vanish $c_{\alpha\alpha}\simeq0$.
This failure could have been anticipated based on the observation
that the tracking does not hold in this case, as observed in \prettyref{fig:symmetric}A.
We therefore need to extend the theory for infinite-sized networks
with homogeneous connectivity.

To this end we write \eqref{eq:corrected_corr_structure} explicitly
for the homogeneous network using $a_{E}=a_{I}=a=(1-m)\, m$. We
see from \eqref{eq:corrected_corr_structure} that $c_{EX}=c_{IX}$
and $c_{EI}=c_{IE}=\frac{1}{2}(c_{EE}+c_{II})$ and introduce $w_{\alpha E}=w$,
$w_{\alpha I}=-gw$, $N_{E}=N_{I}=N$ to obtain

\begin{eqnarray}
\left(2-w(1-g)\right)c_{EX} & = & w\,\frac{a_{X}}{N_{X}}\label{eq:c_ext_symm}\\
c_{EX} & = & \frac{1}{2+w(g-1)}\, w\,\frac{a_{X}}{N_{X}}\nonumber \\
\left[\mathbf{2}-w\left(\begin{array}{cc}
2-g & -g\\
1 & 1-2g
\end{array}\right)\right]\left(\begin{array}{c}
c_{EE}\\
c_{II}
\end{array}\right) & = & 2\frac{w}{N}\, a\,\left(\begin{array}{c}
1\\
-g
\end{array}\right)+2w\, c_{EX}\left(\begin{array}{c}
1\\
1
\end{array}\right)\label{eq:c_int_symm}\\
w & = & K\, J\, S(\mu,\sigma).\nonumber 
\end{eqnarray}
For sufficiently large networks, we can neglect the $2\ll w$ on the
left hand side of \eqref{eq:c_ext_symm} to obtain
\begin{align*}
c_{EX}=c_{IX} & =\frac{1}{g-1}\,\frac{a_{X}}{N_{X}}
\end{align*}
and hence the second equation, again neglecting the $\mathbf{2}\ll w$
on the left hand side, leads to
\begin{align}
\left(\begin{array}{c}
c_{EE}\\
c_{II}
\end{array}\right) & =\left(\begin{array}{c}
c_{EE}^{\mathrm{0}}\\
c_{II}^{0}
\end{array}\right)+\left(\begin{array}{c}
c_{EE}^{1}\\
c_{II}^{\mathrm{1}}
\end{array}\right)\label{eq:cov_symm_infinite}\\
c_{EE}^{0}=c_{II}^{0} & \simeq\frac{1}{g-1}\,\frac{a_{X}}{N_{X}}\nonumber \\
\left(\begin{array}{c}
c_{EE}^{\mathrm{1}}\\
c_{II}^{\mathrm{1}}
\end{array}\right) & \simeq\frac{1}{(g-1)^{2}}\,\left(\begin{array}{c}
-1+2g+g^{2}\\
1+2g-g^{2}
\end{array}\right)\,\frac{a}{N}.\nonumber 
\end{align}
This result shows explicitly the two contributions to the correlations
due to external fluctuations ($c^{0}$) and due to intrinsic fluctuations
($c^{1}$), respectively. In contrast to the case of inhomogeneous
connectivity, both contributions decay as $N^{-1}$, so the external
drive does not provide the leading contribution even in the limit
$N\to\infty$. Note also that we may write this result in a similar
form as for the inhomogeneous connectivity, as 
\begin{align}
c_{\alpha\beta}^{\mathrm{ext.}} & \simeq c_{\alpha\beta}^{0}-\delta_{\alpha\beta}\frac{a_{\alpha}}{N_{\alpha}}\nonumber \\
 & =A\frac{a_{X}}{N_{X}}-\delta_{\alpha\beta}\frac{a_{\alpha}}{N_{\alpha}}\label{eq:c_ext_approx}\\
r_{\alpha\beta}^{\mathrm{int.}} & \simeq c_{\alpha\beta}^{1}+\delta_{\alpha\beta}\frac{a_{\alpha}}{N_{\alpha}}\nonumber \\
\left(\begin{array}{c}
r_{EE}^{\mathrm{int.}}\\
r_{II}^{\mathrm{int.}}
\end{array}\right) & =\frac{2}{(g-1)^{2}}\,\left(\begin{array}{c}
g^{2}\\
1
\end{array}\right)\,\frac{a}{N},\nonumber 
\end{align}
with $A$ given by \eqref{eq:fast_tracking_symm}. Here, $c^{\mathrm{ext.}}\propto N^{-1}$
has the same form as the solution \cite[eqs. (38)-(39)]{Renart10_587}
originating from external fluctuations, but $r^{\mathrm{int.}}\propto N^{-1}$
is still a contribution of same order of magnitude. The susceptibility
$S$ has been eliminated from these expressions and hence only structural
parameters remain, analogous to the solution \cite[eqs. (38)-(39)]{Renart10_587}.
The two contributions $c^{\mathrm{ext.}}=r^{\mathrm{ext.}}-\delta_{\alpha\beta}\frac{a_{\alpha}}{N_{\alpha}}$
and $r^{\mathrm{int.}}$ given by the non-approximate solution of
\prettyref{eq:r_ext} and \prettyref{eq:r_int}, respectively, are
shown together with their sum and with results from direct simulations
in \prettyref{fig:Scaling-to-infinity.}B. For the given network parameters,
the contribution of intrinsic correlations dominates across all network
sizes, because $c_{\alpha\alpha}^{\mathrm{ext.}}\simeq0$, as $A=1$,
and all $N_{\alpha}$ and $a_{\alpha}$ are approximately identical
for $\alpha\in\{E,I,X\}$. The splitting between the covariances of
different types scales proportional to the absolute value $\propto N^{-1}$,
so even at infinite network size the differences between the covariances
stays relatively the same.

The underlying reason for the qualitatively different scaling of the
intrinsically generated correlations $c^{\mathrm{int.}}\propto N^{-1}$
for homogeneous connectivity compared to $c^{\mathrm{int.}}\propto N^{-\frac{3}{2}}$
for inhomogeneous connectivity is related to the one vanishing eigenvalue
of the effective connectivity matrix \eqref{eq:eff_conn_symm}. The
zero eigenvalue belongs to the eigenvector $(g,1)^{T}$, meaning excitation
and inhibition may in this eigendirection fluctuate freely without
sensing any negative feedback through the connectivity, reflected
in the last line in \eqref{eq:c_ext_approx}. These fluctuations are
driven by the intrinsically generated noise of the stochastic update
process and hence contribute notably to the correlations in the network.

In summary, the two examples \nameref{sub:Inhomogeneous-connectivity}
and \nameref{sub:Symmetric-connectivity} are both inhibition-dominated
($g>1$) networks that show small correlations on the order $\frac{a}{N}$
at finite size $N$. Only in the limit of infinitely large networks
with inhomogeneous connectivity is $c^{\mathrm{ext.}}$ the dominant
contribution that can be related to fast and perfect tracking of the
external drive. At finite network sizes, the contribution $c^{\mathrm{int.}}$
is generally not negligible and may be dominant. Therefore fast tracking
cannot be the explanation for small correlations in these networks.
Note that there is a difference in the line of argument used in the
main text of \cite{Renart10_587} and its mathematical supplement:
While the main text advocates fast tracking as the underlying mechanism
explaining small correlations, in the mathematical supplement fast
tracking is found as a consequence of the theory of correlations in
the limit of infinite-sized networks and under the stated prerequisites,
in line with the calculation presented above.

\subsection{Influence of connectivity on the correlation structure\label{sub:conn_struct_corr_struct}}

\begin{figure}
\noindent \centering{}\includegraphics[scale=0.9]{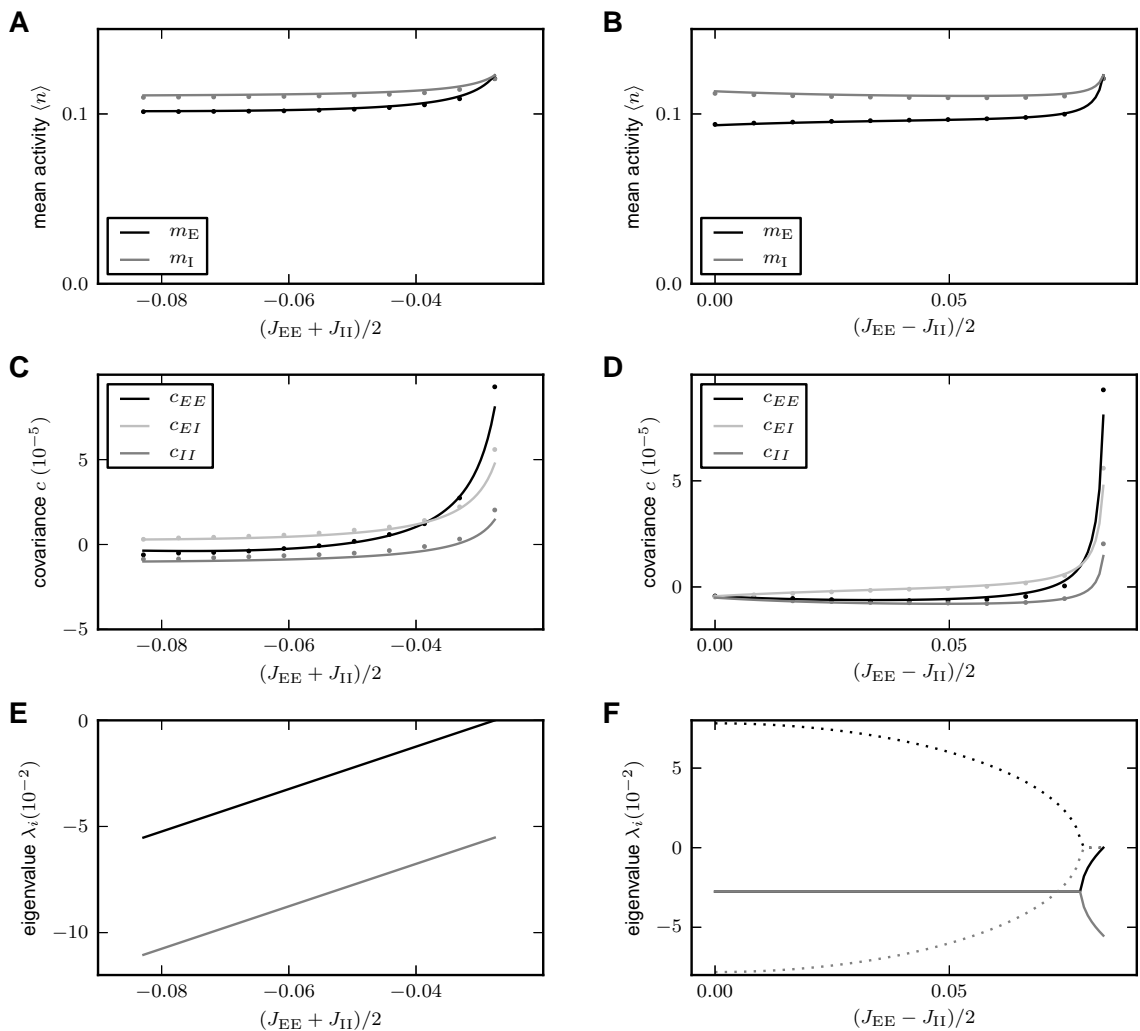}\caption{Connectivity structure determines correlation structure. In the left
column (A,C,E) $c_{1}=(J_{EE}+J_{II})/2$ is the independent variable,
in the right column (B,D,F) $c_{2}=(J_{EE}-J_{II})/2$. \textbf{A},\textbf{B}
Mean activity in the network as a function of the structural parameters
$c_{1}$ and $c_{2}$, respectively. \textbf{C},\textbf{D} Correlations
averaged over pairs of neurons. Dots obtained from direct simulation,
solid curves given by theory \prettyref{eq:corrected_corr_structure}
\textbf{E},\textbf{F} Eigenvalues \eqref{eq:eigenvalues} of the population-averaged
connectivity matrix; solid curves show the real part, dashed curves
the imaginary part.\label{fig:struct_corr}}
\end{figure}

Comparing \prettyref{fig:symmetric}B and \prettyref{fig:renart_binary}B,
the structure of correlations is obviously different. In \prettyref{fig:symmetric}B,
the structure is $c_{EE}>c_{EI}>c_{II}$, whereas in \prettyref{fig:renart_binary}B
the relation is $c_{EI}\simeq c_{EE}>c_{II}$. The only difference
between these two networks are the coupling strengths $J_{II}$ and
$J_{IX}$. In the following we derive a more complete picture of the
determinants of the correlation structure. In order to identify the
parameters that influence the fluctuations in these networks, it is
instructive to study the mean-field equation for the population-averaged
activities. Linearizing \eqref{eq:mean_field_maintext} for small
deviations $\delta n_{\alpha}=n_{\alpha}-m_{\alpha}$ of the population-averaged
activity $n_{\alpha}$ from the fixed point $m_{\alpha}$, for large
networks with $N>K_{\alpha\beta}\gg1$ the dominant term is proportional
to the change of the mean $\delta\mu_{\alpha}=\sum_{\beta}(JK)_{\alpha\beta}\delta n_{\beta}$,
because the standard deviation $\delta\sigma_{\alpha}$ is only proportional
to $\sqrt{K_{\alpha\beta}}$. To linear order we hence have a coupled
set of two differential equations \eqref{eq:mean_field_fluct-1}.
The dynamics of this coupled set of linear differential equations
is determined by the two eigenvalues of the effective connectivity
\begin{eqnarray}
\lambda_{1,2} & = & \eig\left(\{w_{\alpha\beta}\}\right)\nonumber \\
 & = & \frac{w_{EE}+w_{II}}{2}\pm\sqrt{\left(\frac{w_{EE}-w_{II}}{2}\right)^{2}+w_{EI}w_{IE}}.\label{eq:eigenvalues}
\end{eqnarray}
Due to the presence of the leak term on the left hand side of \eqref{eq:mean_field_fluct-1},
the fixed point rate is stable only if the real parts of the eigenvalues
$\lambda_{1,2}$ are both smaller than $1$. In the network with identical
input statistics for all neurons the fluctuating input is characterized
by the same mean and variance $(\mu,\sigma^{2})$ for each neuron.
For homogeneous neuron parameters the susceptibility $S_{\alpha}=S$
is hence the same for both populations $\alpha\in\{E,I\}$. If further
the number of synaptic afferents is the same $K_{\alpha\beta}=K$
for all populations, the eigenvalues can be expressed by those of
the original connectivity matrix as
\begin{eqnarray*}
\frac{\lambda_{1,2}}{S(\mu,\sigma)K} & = & \frac{J_{EE}+J_{II}}{2}\pm\sqrt{\left(\frac{J_{EE}-J_{II}}{2}\right)^{2}+J_{EI}J_{IE}}\\
 & = & c_{1}\pm\sqrt{c_{2}^{2}+J_{EI}J_{IE}},
\end{eqnarray*}
where we defined the two parameters $c_{1}$ and $c_{2}$ which control
the location of the eigenvalues. In the left column of \prettyref{fig:struct_corr}
we keep $J_{EI}$, $J_{IE}$, and $c_{2}$ constant and vary $c_{1}\in[-c_{2},-\sqrt{c_{2}^{2}+J_{EI}J_{IE}}]$,
where we choose the maximum value by the condition $\lambda_{1}<0$
and the minimum value by the condition that $J_{EE}\ge0$ and $J_{II}\le0$,
leading to $c_{1}+c_{2}\ge0$ and $c_{1}-c_{2}\le0$, both fulfilled
if $-c_{2}\le c_{1}\le c_{2}$. Varying $c_{2}$ in the right column
of \prettyref{fig:struct_corr}, the bounds are given by the same
condition that $J_{EE}\ge0$ and $J_{II}\le0$, so $c_{2}\ge0$, and
the condition for the larger eigenvalue to stay below or equal $0$,
so $c_{2}\in[0,\sqrt{c_{1}^{2}-J_{EI}J_{IE}}]$. In order for the
network to maintain similar mean activity, we choose the threshold
of the neurons such that the cancellation condition $0=\sum_{\beta\in\{E,I,X\}}(KJ)_{\alpha\beta}m_{\beta}-\theta$
is fulfilled for $m_{\beta}=0.1$. The resulting average activity
is close to this desired value of $0.1$ and agrees well to the analytical
prediction \eqref{eq:mean_field_maintext}, as shown in \prettyref{fig:struct_corr}A,
B.

The right-most point in both columns of \prettyref{fig:struct_corr}
where one eigenvalue vanishes $\lambda_{1}=0$, results in the same
connectivity structure. This is the case for the connectivity with
the symmetry $J_{EE}=J_{IE}=J$ and $J_{II}=J_{EI}=-gJ$ (cf. \prettyref{fig:symmetric}),
because in this case the population averaged connectivity matrix has
two linearly dependent rows, hence a vanishing determinant and thus
an eigenvalue $0$. As observed in \prettyref{fig:struct_corr}C,D
at this point the absolute magnitude of correlations is largest. This
is intuitively clear as the network has a degree of freedom in the
direction of the eigenvector $v_{1}=(g,1)^{T}$ belonging to the vanishing
eigenvalue $\lambda_{1}=0$. In this direction the system effectively
does not feel any negative feedback, so the evolution is as if the
connectivity would be absent. Fluctuations in this direction hence
become large and are only damped by the exponential relaxation of
the neuronal dynamics, given by the left hand side of \prettyref{eq:mean_field_fluct-1}.
The time constant of these fluctuations is then solely determined
by the time constant of the single neurons, as seen in \prettyref{fig:symmetric}B.
From the coefficients of the eigenvector we can further conclude that
the fluctuations of the excitatory population are stronger by a factor
$g$ than those of the inhibitory population, explaining why $c_{EE}>c_{II}$,
and that both populations fluctuate in-phase, so $c_{EI}>0$, (\prettyref{fig:struct_corr}C,D,
right most point). Moving away from this point, \prettyref{fig:struct_corr}C,D
both show that the magnitude of correlations decreases. Comparing
the temporal structures of \prettyref{fig:symmetric}B and \prettyref{fig:renart_binary}B
shows that also the time scale of fluctuations decreases. The two
structural parameters $c_{1}$ and $c_{2}$ affect the eigenvalues
of the connectivity in a distinct manner. Changing $c_{1}$ merely
shifts the real part of both eigenvalues, but leaves their relative
distance constant, as seen in \prettyref{fig:struct_corr}E. For smaller
values of $c_{1}$ the coupling among excitatory neurons becomes weaker,
so their correlations are reduced. At the left most point in \prettyref{fig:struct_corr}C
the coupling within the excitatory population vanishes, $J_{EE}=0$.
Changing the parameter $c_{2}$ has a qualitatively different effect
on the eigenvalues, as seen in \prettyref{fig:struct_corr}F. At $c_{2}=\sqrt{|J_{EI}J_{IE}|}$,
the two real eigenvalues merge and for smaller $c_{2}$ they turn
into a conjugate complex pair. At the left-most point $J_{EE}-J_{II}=0$,
so both couplings within the populations vanish $J_{EE}=J_{II}=0$.
The system then only has coupling from $E$ to $I$ and vice versa.
The conjugate complex eigenvalues show that the population activity
of the system has oscillatory solutions. This is also called the PING
(pyramidal - inhibitory - gamma) mechanism of oscillations in the
gamma-range \cite{Buzsaki12_203}. \prettyref{fig:struct_corr}C,D
show that for most connectivity structures the correlation structure
is $c_{EI}>c_{EE}>c_{II}$, in contrast to our previous finding \cite{Tetzlaff12_e1002596},
where we studied the symmetric case (the right-most point), at which
the correlation structure is $c_{EE}>c_{EI}>c_{II}$. The comparison
of the direct simulation to the theoretical prediction \prettyref{eq:corrected_corr_structure}
in \prettyref{fig:struct_corr}C,D shows that the theory yields an
accurate prediction of the correlation structure for all connectivity
structures considered here.

\section{Discussion\label{sec:Discussion}}

The present work explains the observed pairwise correlations in a
homogeneous random network of excitatory and inhibitory binary model
neurons driven by an external population of finite size. 

On the methodological side the work is similar to the approach taken
in the work of Renart et al. \cite{Renart10_587}, that starts from
the microscopic Glauber dynamics of binary networks with dense and
strong synaptic coupling $J\propto N^{-\frac{1}{2}}$ and derives
a set of self-consistent equations for the second moment of the fluctuations
in the network. As in the earlier work \cite{Renart10_587}, we take
into account the fluctuations due to the balanced synaptic noise in
the linearization of the neuronal response \cite{Renart10_587,Grytskyy13_258}
rather than relying on noise intrinsic to each neuron, as in the work
by Ginzburg and Sompolinsky \cite{Ginzburg94}. Although the theory
by Ginzburg and Sompolinsky \cite{Ginzburg94} was explicitly derived
for binary networks that are densely, but weakly coupled, i.e. the
number of synapses per neuron is $\propto N$ and synaptic amplitudes
scale as $J\propto N^{-1}$, identical equations result for the case
of strong coupling, where the synaptic amplitudes decay slower than
$N^{-1}$ \cite{Renart10_587}. The reason for both weakly and strongly
coupled networks to be describable by the same theory lies in the
self-regulating property of binary neurons: Their susceptibility (called
$S$ in the present work) inversely scales with the fluctuations in
the input, $S\propto\sigma^{-1}\propto J^{-1}$, such that $JS$ and
hence correlations are independent of synaptic amplitude $J$ \cite{Grytskyy13_258}.
A difference between the work of Ginzburg and Sompolinsky \cite{Ginzburg94}
and the work of Renart et al. \cite{Renart10_587} is, however, that
the first authors assume all correlations to be equally small $\propto N^{-1}$,
whereas the latter show that the distribution of correlations is wider
than their mean due to the variability in the connectivity, in particular
the varying number of common inputs. The theory yields the dominant
contribution to the mean value of this distribution scaling as $N^{-1}$
in the limit of infinite network size. Although the asynchronous state
of densely coupled networks has been described earlier \cite{Vreeswijk96,VanVreeswijk98_1321}
by a mean-field theory neglecting correlations, the main achievement
of the work by Renart et al. \cite{Renart10_587} must be seen as
demonstrating that the formal structure of the theory of correlations
indeed admits a solution with low correlations of order $N^{-1}$
and that such a solution is accompanied by the cancellation of correlations
between the inputs to pairs of neurons. The latter authors employed
an elegant scaling argument, taking the network size and hence the
coupling to infinity, to obtain their results. In contrast, here we
study these networks at finite size and obtain a theoretical prediction
in good agreement with direct simulations in a large range of biologically
relevant networks sizes. We further extend the framework of correlations
in binary networks by an iterative procedure taking into account the
finite-size fluctuations in the mean-field solution to determine the
working point (mean activity) of the network. We find that the iteration
converges to predictions for the covariance with higher accuracy than
the previous method.

Equipped with these methods we investigate a network driven by correlated
input due to shared afferents supplied by an external population.
The analytical expressions for the covariances averaged over pairs
of neurons show that correlations have two components that linearly
superimpose, one caused by intrinsic fluctuations generated within
the local network and one caused by fluctuations due to the external
population. The size $N_{X}$ of the external population controls
the strength of the correlations in the external input. We find that
this external input causes an offset of all pairwise correlations,
which decreases with increasing external population size in proportion
to the strength of the external correlations ($\propto1/N_{X}$).
The structure of correlations within the local network, i.e. the differences
between correlations for pairs of neurons of different types, is mostly
determined by the intrinsically generated fluctuations. These are
proportional to the population-averaged variances $a_{E}$ and $a_{I}$
of the activity of the neurons in the local network. As a result,
the structure of correlations is mostly independent of the external
drive, even for the limiting case of an infinitely large external
population $N_{X}\rightarrow\infty$ or if the external drive is replaced
by a DC signal with the same mean. For the other extreme, when the
size of the external population equals the number of external afferents,
$N_{X}=K$, all neurons receive an exactly identical external signal.
We show that the mechanism of decorrelation \cite{Renart10_587,Tetzlaff12_e1002596}
still holds for these strongly correlated external signals. The resulting
correlation within the network is much smaller than expected given
the amount of common input. In contrast to an earlier explanation
\cite{Renart10_587}, which invokes the network's fast tracking of
the external drive \cite{Vreeswijk96,VanVreeswijk98_1321} as the
cause of small correlations, we here show that the cancellation of
correlations between the inputs to pairs of neurons is equivalent
to a suppression of fluctuations of the population-averaged input
due to negative feedback. This argument is in line with the earlier
explanation that correlations are suppressed by negative feedback
on the population level \cite{Tetzlaff12_e1002596}. Such dominant
negative feedback is a fundamental requirement for the network to
stabilize its activity in the balanced state \cite{Vreeswijk96}.
We further show that the cancellation of input correlations does
not uniquely determine the structure of correlations; different structures
of correlations lead to the same cancellation of correlations between
the summed inputs. The cancellation of input correlations therefore
only constitutes a constraint for the pairwise correlations in the
network. This constraint is trivially fulfilled if the network shows
perfect tracking of external input, which is equivalent to completely
vanishing input fluctuations \cite{Renart10_587}. The correlation
structure in finite-sized networks is in general different from this
limit, but fulfills the constraint imposed by the cancellation of
input correlations.

Performing the limit $N\to\infty$ we distinguish two cases. For an
invertible connectivity matrix, we recover the result by \cite{Renart10_587},
that in the limit of infinite network size correlations are dominated
by tracking of the external signal and intrinsically generated fluctuations
can be neglected; the resulting expressions for the correlations within
the network \cite[Supplementary, eqs. 38,39]{Renart10_587} are lacking
the locally generated fluctuations as additional sources. However,
note that the intermediate result \cite[Supplementary, eqs. 31,33]{Renart10_587}
is identical to \cite[eq. 6.8]{Ginzburg94} and to \eqref{eq:corr_lin}
and contains both contributions.

The convergence of the correlation structure to the limiting theory
appears to be slow. For the parameters given in \cite{Renart10_587},
quantitative agreement is achieved at around $10^{8}$ neurons, which
is beyond the scale up to which random networks are good models for
cortical networks. For the range of biologically relevant network
sizes the correlation structure is dominated by intrinsic fluctuations.
One should note that the lines of argument used in the main text of
\cite{Renart10_587} and in its mathematical supplement are different.
The main text starts at the observation that for an invertible connectivity
matrix and in the inhibition-dominated regime the network activity
exhibits fast-tracking. The authors then argue that hence positive
correlations between excitatory and inhibitory synaptic currents are
responsible for the decorrelation of network activity. The mathematical
supplement, however, first derives the leading term for the pairwise
correlations in the network in the limit of infinite-sized networks
\cite[Supplementary, eqs. 38,39]{Renart10_587} and then shows that
fast tracking and the cancellation of input correlations are both
consequences. For a singular matrix, as for example resulting from
statistically identical inputs to excitatory and inhibitory neurons,
the contributions of external and intrinsic fluctuations both scale
as $N^{-1}$. Hence the intrinsic contribution cannot be neglected
even in the limit $N\to\infty$. At finite network size the observed
structure of correlations generally contains contributions from both
intrinsic and external fluctuations, still present in the intermediate
result \cite[Supplementary, eqs. 31,33]{Renart10_587} and in \cite[eq. 6.8]{Ginzburg94}
and \eqref{eq:corr_lin}. In particular, the external contribution
dominating in infinite networks with invertible connectivity may be
negligible at finite network size. We therefore conclude that the
mechanism determining the correlation structure in finite networks
cannot be deduced from the limit $N\to\infty$ and is not given by
fast tracking of the external signal. Fast tracking is rather a consequence
of negative feedback.

For a common but special choice of network connectivity where the
synaptic weights depend only on the type of the source but not the
target neuron, i.e. $J_{EE}=J_{IE}$ and $J_{EI}=J_{II}$ \cite{Brunel00_183},
we show that the locally generated fluctuations and correlations are
elevated and that the activity only loosely tracks the external input.
The resulting correlation structure is $c_{EE}>c_{EI}>c_{II}$. To
systematically investigate the dependence of the correlation structure
on the network connectivity, it proves useful to parameterize the
structure of the network by two measures differentially controlling
the location of the eigenvalues of the connectivity matrix. We find
that for a wide parameter regime the correlations change quantitatively,
but the correlation structure $c_{EI}>c_{EE}>c_{II}$ remains invariant.
The qualitative comparison with experimental observations of \cite{Gentet10_422}
hence only constrains the connectivity to be within the one or the
other parameter regime.\textbf{}

The networks we study here are balanced networks in the original
sense as introduced in \cite{Vreeswijk96}, that is to say they are
inhibition-dominated and the balance of excitatory and inhibitory
currents on the input side to a neuron arises as a dynamic phenomenon
due to dominance of negative feedback that stabilizes the mean activity.
A network with a balance of excitation and inhibition built into the
connectivity of the network on the other hand would correspond in
our notation to setting $J_{\alpha E}=-J_{\alpha I}$ for both receiving
populations $\alpha\in\{E,I\}$, assuming identical sizes for the
excitatory and the inhibitory population. The network activity is
then no longer stabilized by negative feedback, because the mean activities
$m_{E}$ and $m_{I}$ can freely co-fluctuate, $m_{E}=m_{E}^{0}+\delta m$
and $m_{I}=m_{I}^{0}+\delta m$, without affecting the input to other
cells: $J_{\alpha E}m_{E}+J_{\alpha I}m_{I}$ is independent of $\delta m$.
Mathematically this amounts to a two-fold degenerate vanishing eigenvalue
of the effective connectivity matrix. The resulting strong fluctuations
would have to be treated with different methods than presented here
and would lead to strong correlations. The current work assumes that
fluctuations are sufficiently small so that their effect can be treated
in linear response theory, restricting the expressions to sufficiently
asynchronous and irregular network states. This limitation arises
from the linearization procedure, which approximates the summed synaptic
input by a Gaussian random variable. The deviations of the theory
from direct simulations are stronger at lower mean activity, when
the synaptic input fluctuates in the non-linear part of the effective
transfer function. The best agreement of theory and simulation is
hence obtained for a mean population activity close to $\frac{1}{2}$,
where $1$ means all neurons are active.

For simplicity in most parts of this work we consider networks where
neurons have a fixed in-degree. In large homogeneous random networks
this is often a good approximation, because the mean number of connections
is $pN\propto N$, and its standard deviation $\sqrt{Np(1-p)}\propto\sqrt{N}$
declines relative to the mean. Taking into account distributed synapse
numbers and the resulting distribution of the mean activity in \prettyref{fig:renart_binary}
and \prettyref{fig:Scaling-to-infinity.}A shows that the results
are only marginally affected for low mean activity. The impact of
the activity distribution on the correlation structure is more pronounced
at higher mean activity, where the second moment of the activity distribution
has a notable effect on the population-averaged variance.

The presented work is closely related to our previous work on the
correlation structure in spiking neuronal networks \cite{Tetzlaff12_e1002596}
and indeed was triggered by the review process of the latter. In \cite{Tetzlaff12_e1002596},
we exclusively studied the symmetric connectivity structure, where
excitatory and inhibitory neurons receive the same input on average.
The results are qualitatively the same as those shown in \prettyref{fig:symmetric}.
A difference though is, that the external input in \cite{Tetzlaff12_e1002596}
is uncorrelated, whereas here it originates from a common finite population.
The cancellation condition for input correlations, also observed in
vivo \cite{Okun_535_08}, holds for spiking networks as well as for
the binary networks studied here. For both models, negative feedback
constitutes the essential mechanism underlying the suppression of
fluctuations at the population level. This can be explained by a formal
relationship between both models (see \cite{Grytskyy13_arxiv1304}).

Our theory presents a step towards an understanding of how correlated
neuronal activity in local cortical circuits is shaped by recurrence
and inputs from other cortical and thalamic areas. The correlation
between membrane potentials of pairs of neurons in somatosensory cortex
of behaving mice is dominated by low-frequency oscillations during
quiet wakefulness. If the animal starts whisking, these correlations
significantly decrease, even if the sensory nerve fibers are cut,
suggesting an internal change of brain state \cite{Poulet08_881}.
Our work suggests that such a dynamic reduction of correlation could
come about by modulating the effective negative feedback in the network.
A possible neural implementation is the increase of tonic drive to
inhibitory interneurons. This hypothesis is in line with the observed
faster fluctuations in the whisking state \cite{Poulet08_881}. Further
work is needed to verify if such a mechanism yields a quantitative
explanation of the experimental observations.

The network where the number of incoming external connections per
neuron equals the size of the external population, cf. \prettyref{fig:correlations_ext}
$N_{x}=K$, can be regarded as a setting where all neurons receive
an identical incoming stimulus. The correlations between this signal
and the responses of neurons in the local network (\prettyref{fig:correlations_ext}C)
are smaller than in an unconnected population without local negative
feedback. This can formally be seen from \prettyref{eq:mean_field_fluct-1},
because negative eigenvalues of the recurrent coupling dampen the
population response of the system. This suppression of correlations
between stimulus and local activity hence implies weaker responses
of single neurons to the driving signal. Recent experiments have shown
that only a sparse subset of around 10 percent of the neurons in S1
of behaving mice responds to a sensory stimulus evoked by the active
touch of a whisker with an object \cite{Crochet11_1160}. The subset
of responding cells is determined by those neurons in which the cell
specific combination of activated excitatory and inhibitory conductances
drives the membrane potential above threshold. Our work suggests that
negative feedback mediated among the layer 2/3 pyramidal cells, e.g.
through local interneurons, should effectively reduce their correlated
firing. In a biological network the negative feedback arrives with
a synaptic delay and effectively reduces the low-frequency content
\cite{Tetzlaff12_e1002596}. The response of the local activity is
therefore expected to depend on the spectral properties of the stimulus.
Intuitively one expects responses to better lock to the stimulus for
fast and narrow transients with high-frequency content. Further work
is required to investigate this issue in more detail.

A large number of previous studies on the dynamics of local cortical
networks focuses on the effect of the local connectivity, but ignores
the spatio-temporal structure of external inputs by assuming that
neurons in the local network are independently driven by external
(often Poissonian) sources. Our study shows that the input correlations
of pairs of neurons in the local network are only weakly affected
by additional correlations caused by shared external afferents: Even
for the extreme case where all neurons in the network receive exactly
identical external input ($N_{x}=K$), the input correlations are
small and only slightly larger than those obtained for the case where
neurons receive uncorrelated external input (\foreignlanguage{english}{\textrm{$N_{x}=2N$}};
black curve in \prettyref{fig:struct_corr}C). One may therefore conclude
that the approximation of uncorrelated external input is justified.
In general, this may however be a hasty conclusion. Tiny changes in
synaptic-input correlations have drastic effects, for example, on
the power and reach of extracellular potentials \cite{Linden11_859}.
For the modeling of extracellular potentials, knowledge of the spatio-temporal
structure of inputs from remote areas is crucial.

The theory of correlations in presence of externally impinging signals
is a required building block to study correlation-sensitive synaptic
plasticity \cite{Morrison08_459} in recurrent networks. Understanding
the emerging structure of correlations imposed by an external signal
is the first step in predicting the connectivity patterns resulting
from ongoing synaptic plasticity sensitive to those correlations.

\section*{Acknowledgments\pdfbookmark[1]{Acknowledgments}{AcknowledgmentsPage}}

We thank the two anonymous reviewers for their constructive critique
and in particular for proposing the detailed comparison to \cite{Renart10_587}
with respect to scaling that led to the section ``Limit of infinite
network size''. All simulations were carried out with NEST (http://www.nest-initiative.org).

\end{document}